\documentclass[acmlarge]{acmart}
\usepackage{graphicx} %
\usepackage{natbib}  %
\usepackage{caption} %
\usepackage{algorithm}
\usepackage{listings}
\usepackage{algorithmic}
\usepackage{amsmath}
\usepackage{booktabs}
\usepackage{multirow}
\usepackage[outdir=./]{epstopdf}
\usepackage{enumitem}
\usepackage{caption}
\usepackage{subcaption}
\usepackage{newfloat}
\usepackage{graphicx}
\usepackage{balance}
\usepackage{svg} %
\usepackage[normalem]{ulem}
\usepackage{threeparttable}
\usepackage{booktabs} %

\AtBeginDocument{%
  }

\setcopyright{cc}
\setcctype{by}
\acmJournal{TOIS}
\acmYear{2025} \acmVolume{1} \acmNumber{1} \acmArticle{1} \acmMonth{1} \acmPrice{}\acmDOI{10.1145/3719344}

\begin{document}

\title{Multi-view Intent Learning and Alignment with Large Language Models for Session-based Recommendation}

\author{Shutong Qiao}
\orcid{0000-0002-7368-1535}
\affiliation{
  \institution{School of Big Data and Software Engineering, Chongqing University}
  \city{Chongqing}
  \country{Chongqing, China}
}
\email{qiaoshutong@cqu.edu.cn}

\author{Wei Zhou}
\orcid{0000-0003-0839-8773}
\affiliation{
  \institution{School of Big Data and Software Engineering, Chongqing University}
  \city{Chongqing}
  \country{Chongqing, China}
}
\email{zhouwei@cqu.edu.cn}

\author{Junhao Wen}
\orcid{0000-0002-6561-560X}
\authornote{Corresponding author.}
\affiliation{
  \institution{School of Big Data and Software Engineering, Chongqing University}
  \city{Chongqing}
  \country{Chongqing, China}
}
\email{jhwen@cqu.edu.cn}

\author{Chen Gao}
\orcid{0000-0002-7561-5646}
\authornotemark[1]
\affiliation{
	\institution{BNRist, Tsinghua University}
  \city{Beijing}
  \country{Beijing, China}
}
\email{chgao96@gmail.com}

\author{Qun Luo}
\orcid{0009-0004-8822-4785}
\affiliation{
\institution{Tencent Inc.}
\city{Beijing}
\country{Beijing, China}
}
\email{prisluo@tencent.com}

\author{Peixuan Chen}
\orcid{0009-0006-3508-738X}
\affiliation{
\institution{Tencent Inc.}
\city{Beijing}
\country{Beijing, China}
}
\email{noahchen@tencent.com}

\author{Yong Li}
\orcid{0000-0001-5617-1659}
\affiliation{
	\institution{Department of Electronic Engineering, BNRist, Tsinghua University}
	\city{Beijing}
	\country{Beijing, China}
}
\email{liyong07@tsinghua.edu.cn}

\renewcommand{\shortauthors}{Shutong Qiao et.al.}

\begin{abstract}
Session-based recommendation (SBR) methods often rely on user behavior data, which can struggle with the sparsity of session data, limiting performance. Researchers have identified that beyond behavioral signals, rich semantic information in item descriptions is crucial for capturing hidden user intent. While large language models (LLMs) offer new ways to leverage this semantic data, the challenges of session anonymity, short-sequence nature, and high LLM training costs have hindered the development of a lightweight, efficient LLM framework for SBR.

To address the above challenges, we propose an LLM-enhanced SBR framework that integrates semantic and behavioral signals from multiple views. This two-stage framework leverages the strengths of both LLMs and traditional SBR models while minimizing training costs. In the first stage, we use multi-view prompts to infer latent user intentions at the session semantic level, supported by an intent localization module to alleviate LLM hallucinations. In the second stage, we align and unify these semantic inferences with behavioral representations, effectively merging insights from both large and small models.
Extensive experiments on two real datasets demonstrate that the LLM4SBR framework can effectively improve model performance.
We release our codes along with the baselines at \url{https://github.com/tsinghua-fib-lab/LLM4SBR}.
\end{abstract}



\maketitle

\section{Introduction}
\label{sec::intro}
With the widespread application of Recommender systems (RS) \cite{lu2015recommender, he2017neural, yuan2024fellas, yuan2024ptf} in multiple fields, traditional RS methods are no longer sufficient to meet users' actual needs in real-world scenarios. Specifically, on the one hand, for the new-coming user, it is difficult to collect enough user behavior data and preference information to make accurate recommendations and encounter the "cold start" problem. On the other hand, with the enhancement of awareness of data privacy protection globally, such as California's California Consumer Privacy Act (CCPA) \cite{illman2019california}, which reinforces the protection of individual privacy rights, placing compliance pressure on traditional recommendation strategies that rely on user personal information. Therefore, against this backdrop, research on Session-based Recommendation (SBR) \cite{wu2019session} technology is gradually becoming a hotspot, which aims to provide recommendations based on users' transient anonymous session behavioral data with less risk to user privacy.

Traditional research in SBR relies heavily on deep neural networks to model user interaction sequences. This includes the use of Recurrent Neural Networks (RNNs) \cite{mandic2001recurrent} with Attention Mechanisms to capture long-term dependencies between items in a session, as well as the use of Graph Neural Networks (GNNs) \cite{scarselli2008graph} and Hypergraph Neural Networks (HGNNs) \cite{feng2019hypergraph} to model complex item transformation relationships in a session. 
Although these deep learning-based SBR models \cite{adomavicius2005toward,rendle2010factorizing,hidasi2015session,wu2019session,wang2019collaborative} show high effectiveness in modeling behavioral information, they are still faced with the challenge of high sparsity of session data. Specifically, the SBR model uses one-hot encoding to represent item IDs, greatly weakening the correlation between items. Therefore, it is difficult for the model to deeply analyze and grasp the user's fundamental interests and potential intentions when only analyzing sparse behavior records. 
On the other hand, unlike interactive data, semantic data revolves around items' inherent similarity and relevance, which may bring opportunities for the above challenge.
For instance, assuming a user sequentially interacts with ``\textit{iPhone 15, running shoes, iPhone 14, milk, skirt}", traditional SBR models may interpret recent interactions with ``\textit{milk}" and ``\textit{skirt}" as indicative of the user's prevailing preferences. 
However, through a deeper semantic lens, it becomes apparent that the selection of ``\textit{iPhone 15}'' and ``\textit{iPhone 14}'' underscores a more enduring and profound interest in Apple-branded products. 
This underscores the potential of semantic information integration within RSs to more accurately discern and capture the underlying needs and users' interests.

Currently, the rapid development of large language models (LLMs) \cite{achiam2023gpt,touvron2023llama,zeng2022glm, qwen} has made remarkable achievements in the depth and breadth of language understanding and shows a strong logical inference ability, opening up a new path for recommendation algorithms. Recent work \cite{cui2022m6, geng2022recommendation} has begun to utilize pre-training and fine-tuning techniques to integrate core tasks in recommendation systems, transforming them into language understanding and generation problems in natural language processing (NLP). Through this approach, LLMs can generate personalized recommendation lists based on a comprehensive understanding of users' past behaviors and personal preferences, providing detailed recommendations and significantly enhancing both user experience and recommendation accuracy. However, the high performance of LLM is often accompanied by high computational costs and latency issues, which largely limits their widespread application in real-time industrial scenarios. Therefore, ID-based collaborative information modeling methods still maintain their unique advantages and indispensable position.
In addition, due to the high sparsity and anonymity of session data, research on incorporating LLM into SBR scenarios is still in its infancy. Current research mainly utilizes the capabilities of LLM by fine-tuning parameters\cite{bao2023bi, bao2023tallrec} or prompt optimization\cite{sun2023large}.
The above work has conducted a preliminary exploration, with unresolved key challenges as follows:
\begin{itemize}[leftmargin=*]
    \item Behavioral information based on the ID paradigm is the core of personalized recommender systems, which is even ignored by some existing methods solely based on semantic information.
    \item Session data is often augmented using sequence segmentation, but this approach may result in a large number of similar session sequences in the generated dataset, which affects the fine-tuned LLMs based on this data to produce repetitive sentences when generating responses.
    \item Fine-tuning and pre-training often require a large number of computational resources, which is difficult to deploy in real-world scenarios practically. 
    \item The high sparsity of the session data makes LLMs fail to generate valid answers or may generate incorrect items that are outside the candidate set.
\end{itemize}

To address these challenges, we propose a lightweight and effective LLM-enhanced framework (LLM4SBR), which comprises two distinct stages: intent inference and representation enhancement. 
In the intent inference stage, we employ LLM as the inference engine by guiding LLM to infer through carefully designed prompts from different views of user intents. The intent localization module is crafted to alleviate hallucinations and semantically enhance the inference results.
In the representation enhancement stage, we introduce the traditional SBR model to load interaction behaviors simultaneously, combined with textual intention inference results. On the one hand, traditional models generate conversation representations from different views based on interaction data, while on the other hand, it parses text data into embedded forms. Subsequently, each view performs alignment and uniformity of session and inference embeddings separately. Finally, we fuse all embeddings from all views as the session representation for prediction.

We summarize the contributions of this work as follows:
\begin{itemize}[noitemsep, topsep=2pt, leftmargin=*]
    \item We take the pioneering step to propose a general framework for introducing large language models to the session-based recommendation, aiming to leverage LLM's powerful logical-reasoning capability to facilitate deep semantic information integration. The framework ingeniously separates LLM's inference process from the training of the traditional SBR model, forming a two-stage strategy, thereby achieving a balanced optimization of efficiency and effectiveness.
    
    \item We first propose an intent localization module, which can alleviate LLM hallucination and enhance semantic-level intent in preliminary results of LLM inference. We then achieve a finer-grained modal alignment by performing alignment and uniformity between embeddings from different views, facilitating the effective integration of interaction ID information and semantic information.

    \item We conduct extensive experiments on two real-world datasets, and the results show that our proposed LLMSBR framework can be a plug-in solution that can steadily improve the performance of almost all the existing SBR models. Further studies verify the effectiveness of each component in our proposed LLMSBR framework.
\end{itemize}

\section{Related Work}
\label{sec::related}
\subsection{Session-based Recommendation}
\label{sec::relatedwork1}
The available information in the field of SBR is very limited, consisting only of interaction data within the session. Therefore, SBR research focuses on how to effectively model interaction behavior and learn session preferences. Based on different modeling emphases, we can broadly categorize SBR methods into two types: traditional SBR methods and methods focusing on modeling item transition relationships.

In traditional SBR methods, S-POP \cite{adomavicius2005toward} recommends based on the most popular items, and Item-KNN \cite{davidson2010youtube} calculates item similarity based on historical behavior to recommend similar items. As Markov chains exhibit advantages in modeling sequential data, FPMC \cite{rendle2010factorizing} captures data sequence information and user preferences by combining first-order Markov chains with matrix factorization. In the SBR methods based on deep learning, inspired by the field of NLP, GRU4Rec \cite{hidasi2015session} proposed for the first time to use of the RNN to simulate user preference changes in behavioral sequence data. Based on this research, Stamp improved performance by introducing an attention mechanism to make preferences more targeted. NARM \cite{li2017neural} uses the attention network to capture users' short-term interests and long-term dependencies. DSAN\cite{yuan2021dual} adaptively filters the noisy information in the sequences through a two-layer sparse attention network, ensuring that the model can focus on items that truly reflect the user's interests.

As GNNs show their prowess in various fields, SBR researchers have found that by constructing session data into the form of graphs, they can better capture the complex transformation relationships between items and greatly improve recommendation performance.
SR-GNN\cite{wu2019session} is the first model to represent sequences in the form of session graphs, utilizing gated graph neural networks as encoders. GC-SAN \cite{xu2019graph}, an upgraded version of SR-GNN, incorporates attention mechanisms to make session representations more targeted.
GCE-GNN \cite{wang2020global}, MSGAT \cite{qiao2023bi}, MEGAN \cite{wang2023multi} and KMVG\cite{chen2023knowledge} construct multiple graphs with different structures, simultaneously considering both local item collaborations and global session collaboration relationships.
In addition, DHCN \cite{xia2021self}, HL\cite{wang2022exploiting}, and HIDE \cite{li2022enhancing} captures the complex high-order miscellaneous information of the items by building the hypergraph. In recent research, GSNIR \cite{jin2023dual} adopts a dual intent network to learn user intent from the attention mechanism and historical data distribution respectively. MiaSRec \cite{choi2024multi} represents various session intents by deriving multiple session representations centered around each item and dynamically selecting important representations.

Although the aforementioned SBR methods have achieved good performance, they mainly focus on modeling the interaction information in the session but fail to fully explore and utilize the rich semantic information contained in the sequence.

\subsection{Recommender System with LLM}
\label{sec::relatedwork2}
Generative dialogue models represented by ChatGPT have sparked research in various fields. According to how LLM participates in the RS, we divide it into LLM as Recommender and LLM-enhanced Recommender.

\subsubsection{LLM as Recommender}
The model of LLM as Recommender realizes the transformation from the ID paradigm to the modal paradigm by converting the recommendation task into a task in natural language processing. The M6-Rec \cite{cui2022m6} model extends the pre-trained language model M6 \cite{lin2021m6} by transforming recommendation tasks into either language understanding or language generation tasks. It establishes a unified foundational recommendation model to reduce downstream tasks' dependence on data. Geng et al. \cite{geng2022recommendation} proposed the P5 paradigm, which enables predictions in a zero-shot or few-shot manner by providing adaptive personalized prompts tailored to different users. This approach reduces the need for fine-tuning. Kang et al. \cite{kang2023llms} evaluated the performance of LLMs of different sizes (250M - 540B parameters) in zero-shot, few-shot, and fine-tuning scenarios to explore the extent to which LLM understands user preferences based on the user's previous behavior. Sunhao Dai et al. \cite{dai2023uncovering} enhance the recommendation capabilities of ChatGPT by combining ChatGPT with traditional information retrieval (IR) ranking functions. GPT4Rec \cite{li2023gpt4rec} first generates queries based on a language model, and then optimizes product retrieval separately through a search engine, addressing optimization from two aspects. VIP5 \cite{geng2023vip5} explores a multi-modal base model of the P5 recommendation paradigm that considers both visual and textual modalities. Zhu Sun et al. \cite{sun2023large} proposed the PO4ISR model of SBR, which promotes LLM to continuously reflect and update the results from the view of real-time optimization prompts to improve the accuracy of recommendations. Agent4Rec \cite{zhang2023generative} utilizes a generative agent empowered by LLM to simulate and infer personalized user preferences and behavioral patterns. ToolRec \cite{zhao2024let} uses LLM as a proxy user to guide the recommendation process and call external tools to generate recommendation lists that are closely related to the user's subtle preferences. Lin et al. \cite{lin2024data} achieve the goal of efficiently fine-tuning LLM-based RS by introducing a data-pruning task. 

Although these methods have made breakthrough progress in zero-shot, few-shot, and interpretability aspects, the core of the above method is to enhance recommendation performance by improving LLM's adaptability to recommended data and inference capabilities. Therefore, they suffer from drawbacks such as high fine-tuning costs and difficulty capturing specific fine-grained behavioral patterns.

\subsubsection{LLM-enhanced Recommender}
LLM-enhanced RS treats LLM as a tool to enhance the performance of recommendation models. The large-model recommendation framework proposed by Weiwei et al. \cite{wei2024llmrec} leverages graph-enhanced strategies based on LLM to enhance RS, addressing challenges posed by data sparsity and low-quality side information in RS. Chat-Rec \cite{gao2023chat} integrates traditional RS with conversational AI like ChatGPT, eliminating the need for training to gain a deep understanding of user preferences through LLM's comprehension of dialogue context. CTRL \cite{li2023ctrl} regards the original table data and the corresponding text data as two different modalities, uses the collaborative CTR model and the pre-trained language model, respectively, for feature extraction, and adjusts the knowledge of the two modalities through comparative learning. E4SRec \cite{li2023e4srec} is a solution that combines sequence recommendation with LLMs. It takes only ID sequences as input and ensures efficient controllable generation by predicting all candidate sequences at each forward pass.
The above method has made us realize the potential of integrating LLM with RS and how a two-stage framework can better balance efficiency and performance compared to an end-to-end framework. 
Jesse Harte et al. \cite{harte2023leveraging} devised three strategies for leveraging LLM, and found that using embeddings initialized with LLM significantly enhances the performance of sequential recommendation models. This inspires us about the importance of textual semantics. 
The SAID \cite{hu2024enhancing} shows how to effectively use LLM to convert item IDs into semantically rich embedding vectors for downstream recommendation task models.

The above methods explore the effectiveness of LLM in RS from different views, allowing us to realize the potential of integrating LLM with RS and how the two-stage framework can better balance efficiency and performance compared with the end-to-end framework. However, these methods fail to fully integrate and fuse semantic information and behavioral data, and therefore perform poorly when dealing with SBR scenarios with high sparsity and lack of user information.

\begin{table}[htbp]
\centering
\caption{Description of notations.}
\label{tab:symbol_definitions}
\begin{tabular}{cp{10cm}} %
\toprule
\textbf{Notations} & \textbf{Descriptions} \\
\midrule
$S$  &   The set of all session sequences \\
$s_t$  &  The session sequence of clicks at timestamp $t$  \\
$I$ &  The set of all items  \\
$i_{t,k}$ & The $k \textit{-}$th item clicked in the session at timestamp $t$   \\
$Y$, $y$, $\hat{y}$ &  The label set of all sessions, the ground truth and predicted label of a single session  \\
$\text{Text}_{\text{infer}}$, $\text{Text}_{\text{item}}$ & LLM inference results and item names in text form \\
$E$, $e$ &  Text embeddings obtained using a pre-trained language model \\
$H$, $h$ &  The set of latent representations for all sessions and the latent representation for a single session\\
$d$,$W$,$b$ &  Dimension, weight matrix and biases  \\
$T$  & Alignment transformation matrix \\
$K$ & The number of items that users are actually interested in \\
$r$ & The number of items with similar semantics\\
$\tau$ & Weight of alignment and uniformity loss functions \\
$\mathcal{L}$ &  Loss Function\\
\bottomrule
\end{tabular}
\end{table}

\section{Problem Formulation}
The objective of SBR is to predict the next interaction item expected to occur in the current session history of an anonymous user. Here, we provide the problem definition in mathematical terms.
Each data entry in the dataset represents a session sequence. Let the collection of all sessions be denoted as $\mathcal{S}=\{s_1, s_2,\cdots, s_m\}$, where $m$ is the total number of sessions.
The item set is the summary of items that have appeared in all sessions, which we define as $\mathcal{I} = \{i_1, i_2, \cdots, i_n\}$, where $n$ is the total number of items in the set.
We represent the t-th session $s_t$ in the dataset as $s_t = \{i_{t,1},i_{t,2},\cdots,i_{t,k},\cdots, i_{t,|s_t|}\}$, where $|s_t|$ is the length of the current session, and $i_{t,k} \in \mathcal{I}$ represents the $k$-th clicked item in the current session $s_t$. 
Based on the above symbols and descriptions, we define the modeling goal of session $s_t$ as predicting the click of the $|s_t|+1$th item based on the historical behavior records of $s_t$. The symbols and explanations involved in this article are listed in Table \ref{tab:symbol_definitions}. In the following text, different symbols are distinguished by subscripts.

\section{METHODOLOGY}
\label{sec::method}
\begin{figure*}[t]
    \centering
    \includegraphics[width=1.0\textwidth]{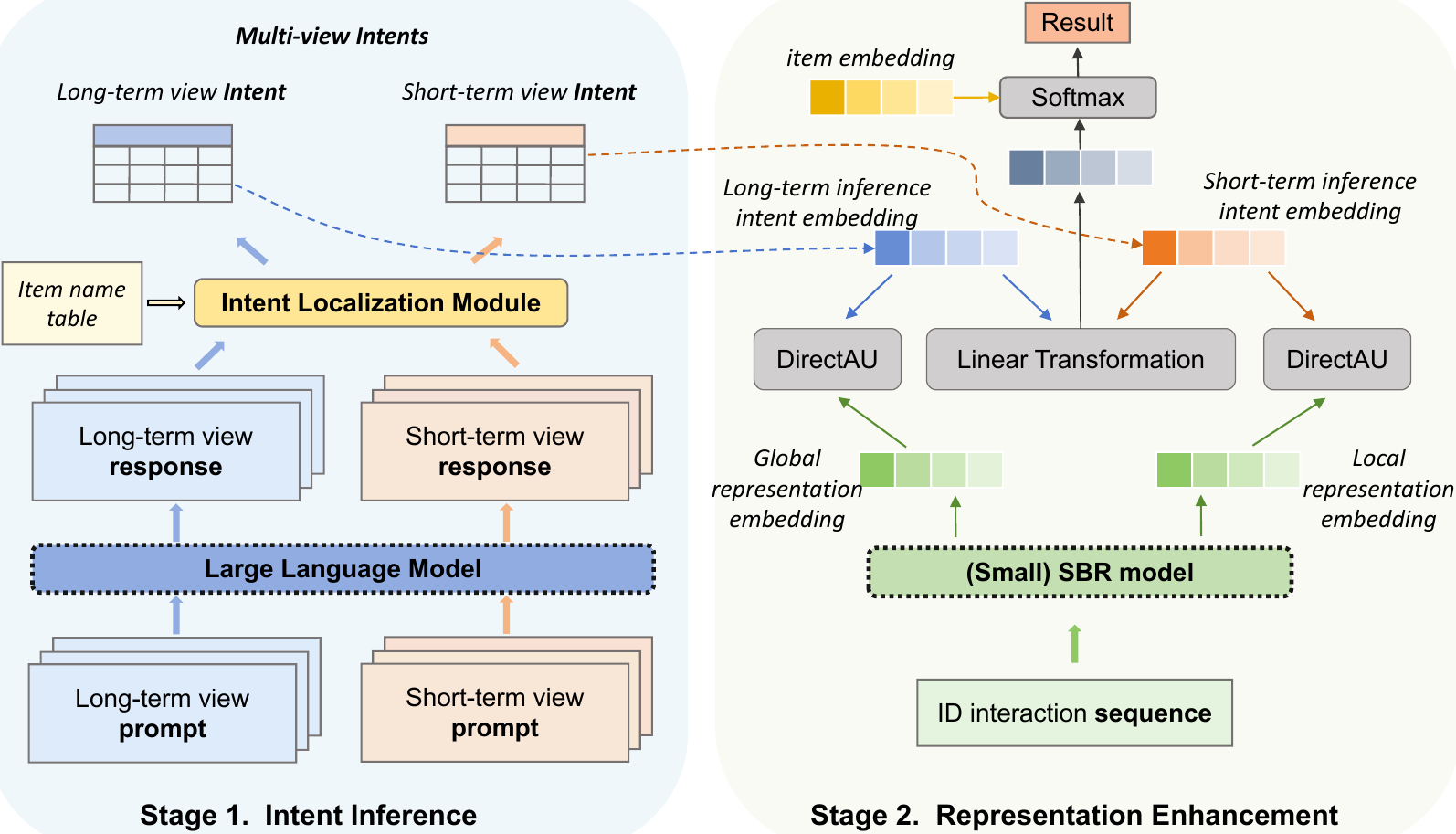}
    
    \caption{LLM4SBR framework diagram. LLM4SBR is a two-stage framework: (a) In the intent inference stage, LLM makes initial inferences based on prompts from different views (long-term and short-term). Subsequently, the intent localization module is utilized to alleviate hallucinations and enhance semantics in the inference results. (b) In the representation enhancement stage, interaction data and text data are synchronously loaded into the model. Traditional SBR models are used to model the interaction data to obtain local and global session representations. After aligning and uniforming session representations and inference representations of the same view, all representations are fused into the final session representation for prediction.} 
    \label{fig:1}
\end{figure*}

The overall architecture of LLM4SBR is depicted in Figure \ref{fig:1}, and the framework process is shown in Algorithm \ref{alg:LLM4SBR}. This section will introduce the problem definition and the specific implementation details of each module in turn.

\subsection{Intent Inference Stage}
\subsubsection{Prompt Design}
To more effectively utilize the inference capabilities of LLM and better integrate the inferential results of LLM with the representations learned by the SBR model, we introduce view constraint qualifiers as an auxiliary tool. Specifically, in our prompt design, we utilize the view-limiting qualifiers based on commonly used behavioral modeling views in SBR (long-term and short-term). 
By artificially setting them, we decompose the text inference task into finer-grained view inference subtasks, thereby maximizing the utilization of LLM's inference capabilities. It is worth noting that the view settings are not fixed and can be freely added or removed, endowing the framework with scalability.

\begin{figure}[t]
    \centering
    \includegraphics[width=0.8\textwidth]{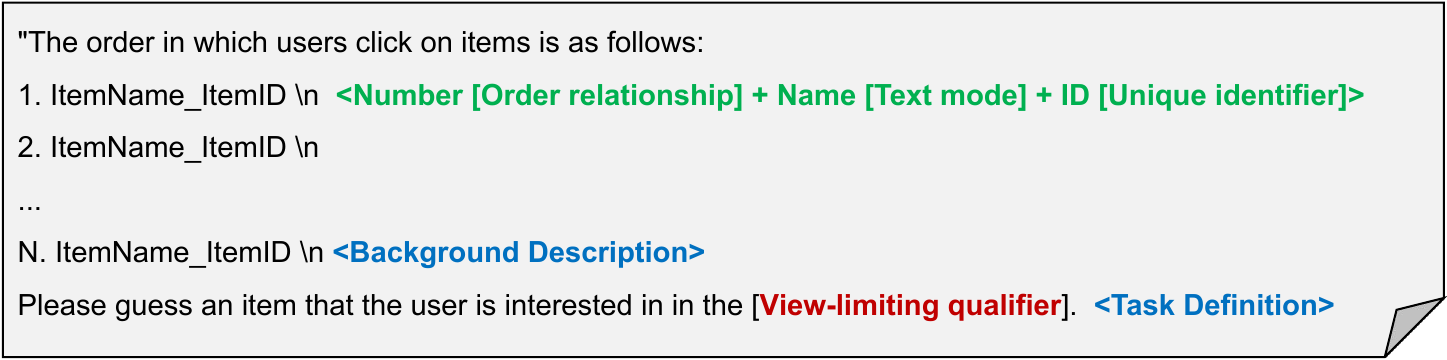}
    \caption{Illustration of the design of prompts. LLM will perform inference based on short-term and long-term prompts respectively to obtain inference results from different views.} 
    \label{fig:template}
\end{figure}

The specific prompt template is shown in figure \ref{fig:template}, where we mark the template's necessary components in blue, the view-limiting qualifier in red, and the composition of the sequence items in green. A prompt consists of two parts: [\textbf{Background Description} and \textbf{Task Definition}]. 
Each item in the sequence of the background description consists of three components: $<\text{Number}+\text{Name}+\text{ID}>$. The "\textbf{Number}" indicates the order of the items, while the "\textbf{Name}" represents the textual modality information. 
Some studies \cite{li2023prompt, hua2023index} suggest that ID information helps LLM distinguish between different items more accurately. Inspired by this, we incorporate corresponding ID information after the item names in the prompt design. 

\subsubsection{LLM Inference}
\begin{figure}[t]
    \centering
    \includegraphics[width=0.92\textwidth]{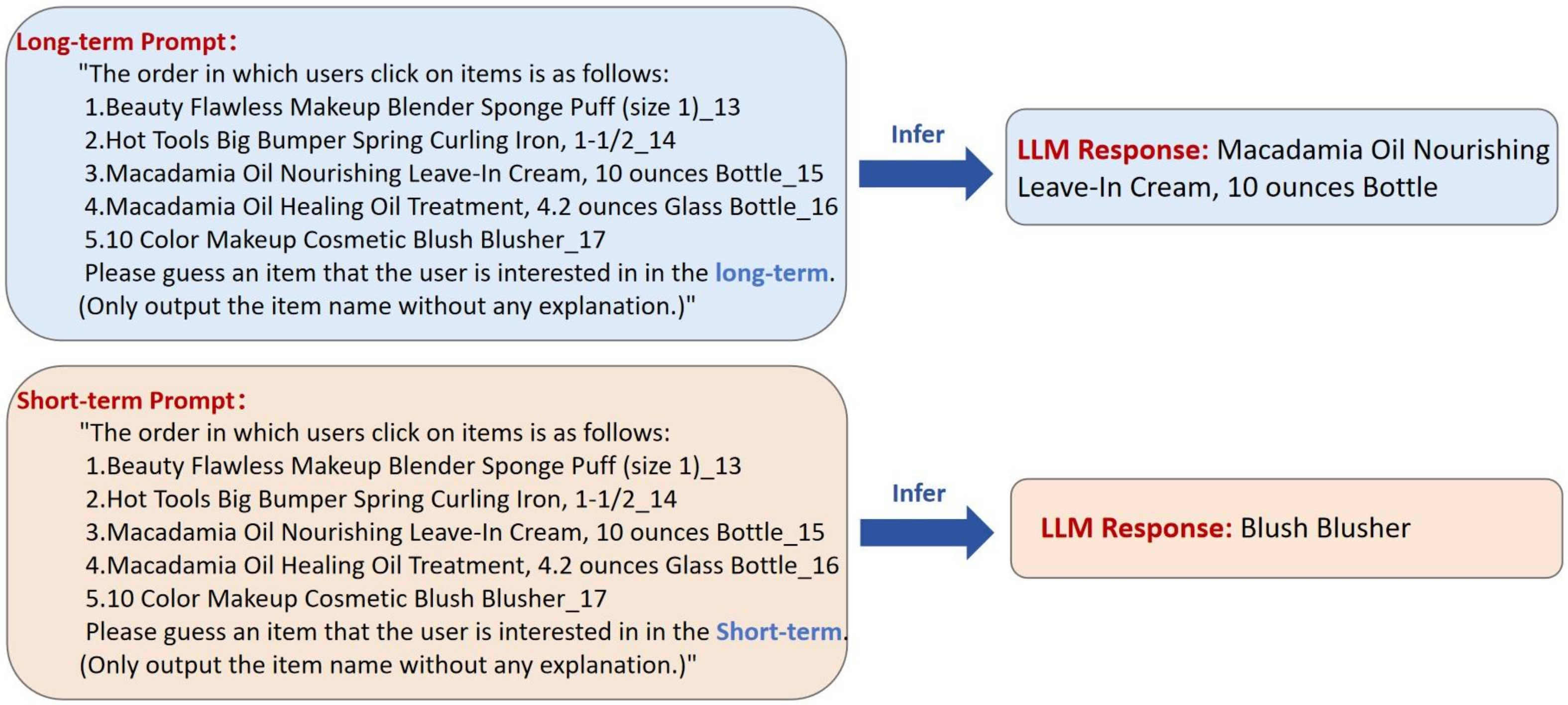}
    \caption{An example of LLM inference from different views.} 
    \label{fig:example}
\end{figure}
To enhance the effective utilization of semantic information and understand the genuine intent of sessions, we leverage the contextual understanding and logical inference capabilities of LLMs to achieve intent inference from different views. Figure \ref{fig:example} shows an example of LLM reasoning based on prompts from different views. 
It is worth noting that the LLM is interchangeable here. LLMs with more parameters and stronger inference capabilities can produce more accurate inference results. We adopt the question-and-answer format, input different view prompts as questions to the LLM, and then the LLM returns its inferring results according to the prompts. 
\begin{equation}
    \text{Text}_{\text{infer}} = \textbf{\text{LLM}}(\text{prompt}_{\text{p}})
\end{equation}

\subsubsection{Intent Localization}
\begin{figure}[t]
    \centering
    \includegraphics[width=0.6\textwidth]{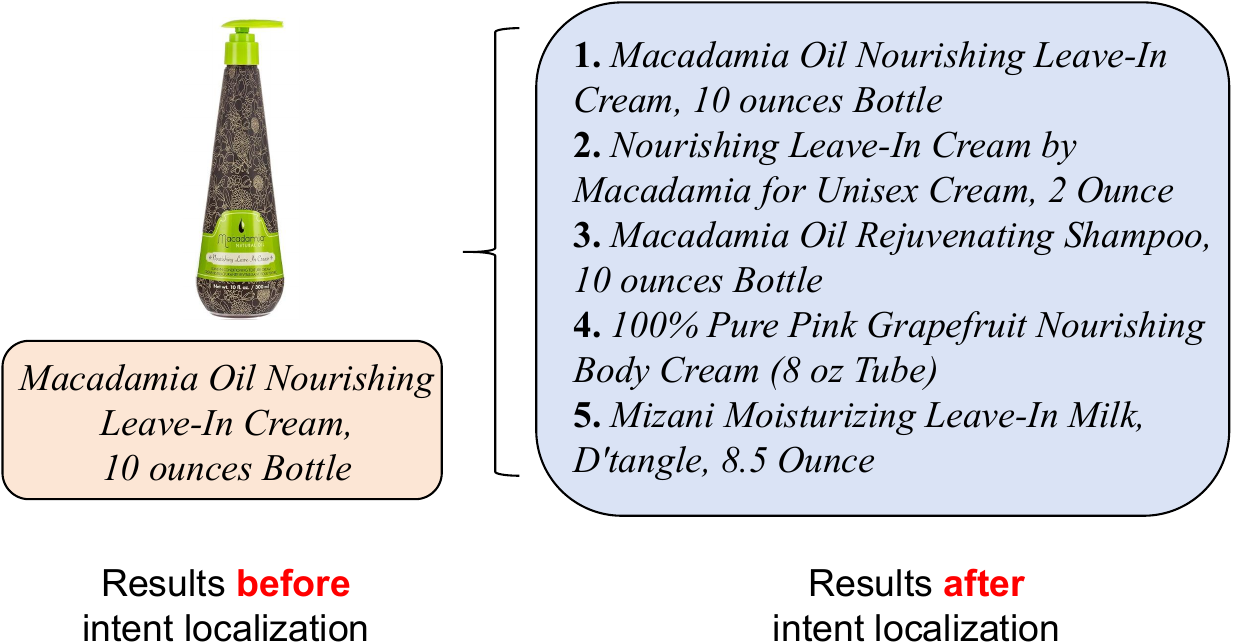}
    
    \caption{The result of intent localization module.} 
    \label{fig:intent localization}
\end{figure}

To assist LLM in alleviating hallucinations and achieving semantic enhancement, we designed the intent localization module.
Although in most cases, the LLM inference result is an accurate item name, sometimes it may be just a vague item category or key item term. In rare cases, a reasonable inference result may not be obtained. The red portion in Figure \ref{fig:intent localization} illustrates the initial inference results of LLM.

Inspired by the RAG retrieval model \cite{lewis2020retrieval}, alleviating hallucinations in LLM requires providing relevant external knowledge to LLM. The text retrieval scheme of the RAG model is usually based on the similarity of text embeddings, so we first encode all inference results and the text of the item set into embedding forms using a pre-trained BERT model \cite{devlin2018bert}\footnote{https://huggingface.co/bert-base-uncased}.
\begin{equation}
    E_{\text{infer}} = \textbf{\text{Bert}}(\text{Text}_{\text{infer}}),
\end{equation}
\begin{equation}
    E_{\text{item}} = \textbf{\text{Bert}}(\text{Text}_{\text{item}}),
\end{equation}
where $e_{\text{infer}}, e_{\text{item}} \in \mathbb{R}^{d_{\text{text}}}$.

Then, we compute the cosine similarity scores between each inference result and all item embeddings. Utilizing text embedding similarity, we select the Top-$r$ most similar actual items from the item set, where $r$ is a hyperparameter that controls the number of semantically similar items to be filtered. We multiply the embeddings of selected items by their corresponding similarity scores and then sum them up to obtain the inference result of the LLM. Figure \ref{fig:intent localization} illustrates the comparison of inference results before and after using the intent localization module. Finally, the inference results of each view are passed through this module to alleviate hallucinations and enhance semantics.
\begin{equation}
    \text{Similarity}^i = \frac{{e_{\text{infer}}^i} \cdot e_{\text{item}}^j}{\|e_{\text{infer}}^i\| \|e_{\text{item}}^j\|}, \label{eq:Simi}
\end{equation}

\begin{equation}
    h_{\text{infer}} = \sum_{i_r \in I_r} \text{Similarity}^{i_r} e_{\text{item}}^{i_r} , \label{eq:filter}
\end{equation}
where $e_{\text{infer}}^i \in E_{\text{infer}}$ is the text embedding of the inference result and $e_{\text{item}}^j \in E_{\text{item}}$ is the text embedding of the item name. The formula \eqref{eq:Simi} represents the calculation of the cosine similarity scores between each inference result and all item names. $I_r$ denotes the set of the Top-$r$ item indices with the highest similarity scores, in which $r$ is a hyperparameter.

\begin{algorithm}[t]
    \caption{Process details of the LLM4SBR framework.}
    \label{alg:LLM4SBR}
    \begin{algorithmic}[1]
        \REQUIRE prompts(Short $\&$ Long views), $I_t$, $\text{Text}_{\text{item}}, $
        \ENSURE $\hat{y_i}$
        \STATE $E_{\text{item}} = \textbf{BERT}(\text{Text}_{\text{item}})$
        \FOR{each prompt $p \in \{P_{short}, P_{long}\}$}
            \STATE $\text{Text}_{\text{infer}} = \textbf{LLM}(p)$
            \STATE $E_{\text{infer}} = \textbf{BERT}(\text{Text}_{\text{infer}})$
            \STATE $\text{Similarity}^i = \frac{{e_{\text{infer}}^i} \cdot e_{\text{item}}^j}{\|e_{\text{infer}}^i\| \|e_{\text{item}}^j\|}$
            \STATE $h_{\text{infer}} = \sum_{i_r \in I_r} \text{Similarity}^{i_r} e_{\text{item}}^{i_r}$, where $I_r$ is the set of the most similar $r$ items.
        \ENDFOR
        \STATE $H_t^l, H_t^g = \textbf{\text{SBR}\textit{-}\text{Model}}(\mathcal{I}_t)$
        \STATE Initial $\mathcal{L}_{a} = 0, \mathcal{L}_{u} = 0$
        \FOR{each views $\text{p}$ (long-term \& short-term)}
        \STATE $\Tilde{h_{\text{infer}}^\text{p}} = \mathbf{T} h_{\text{infer}}^\text{p}$
            \STATE 
    $\mathcal{L}_{a}^{\text{p}} = \underset{(\text{infer},\text{t}) \sim sess}{\mathbb{E}}||\Tilde{h_{\text{infer}}^\text{p}} - h_\text{t}^\text{p} ||^2$
            \STATE
    $\mathcal{L}_{u}^{\text{p}} = {\log e^{-2 {||\Tilde{h_{\text{infer}}^\text{p}} - \Tilde{h_{\text{infer}^{'}}^\text{p}}||}^{2}}/2} + {\log e^{-2 {||h_{t}^\text{p} - h_{t^{'}}^\text{p}||}^{2}}/2} $
        \STATE $\mathcal{L}_{a} += \mathcal{L}_{a}^{\text{p}}$, $\mathcal{L}_{u} += \mathcal{L}_{u}^{\text{p}}$
        \ENDFOR
        \STATE $\mathcal{L}_{a} = \Bar{\mathcal{L}_{a}}, \mathcal{L}_{u} = \Bar{\mathcal{L}_{u}}$
        \STATE $
        \alpha_{\text{infer}}^{\text{lt}} = \textbf{Q}_1^\mathrm{T}(\text{sigmoid}(H_\text{t}^{g} + H_{\text{infer}}^{\text{lt}}))
        $
        \STATE $
        \alpha_{\text{infer}}^{\text{st}} = \textbf{Q}_2^\mathrm{T}(\text{sigmoid}(H_\text{t}^{l} + H_{\text{infer}}^{\text{st}}))$
        \STATE $
        H_{\text{sess}} = \textbf{W} [ H_\text{t}^{l} * \alpha_{\text{infer}}^{\text{st}} ; H_\text{t}^{g} * \alpha_{\text{infer}}^{\text{lt}}], \label{conbine}
        $
        
        \STATE $\hat{y_i} = \text{softmax}(h_{\text{sess}}^\mathrm{T} v_{i})$
        \STATE $\mathcal{L}_{r} = - \sum_{i=1}^n y_i log(\hat{y_i})+(1-y_i)log(1-\hat{y_i})$
        \STATE $\mathcal{L} = \mathcal{L}_{r} + \tau \left( {L}_{a} + {L}_{u} \right)$
        \RETURN $\hat{y_i}$
    \end{algorithmic}
\end{algorithm}

\subsection{Representation Enhancement Stage}
After the intent inference stage, we move into the representation enhancement phase. In this stage, the SBR model processes behavioral modeling data and parsed inference data. Subsequently, the alignment and uniformity of session embeddings and inference embeddings are conducted separately for each view. Ultimately, all view inference embeddings are fused with session embeddings to form the final session representation used for prediction.

Most of the state-of-the-art (SOTA) models in RS are currently based on the item-ID paradigm. Although this paradigm may sacrifice semantic information, its performance and efficiency are undeniably superior. There is still a long way to go to subvert the ID paradigm. \cite{yuan2023go} Therefore, we opt to model user behavior based on the item-ID paradigm while simultaneously injecting multimodal information for supplementary enhancement. The SBR model in the framework is interchangeable. In the subsequent experimental section, we also test the performance after replacing SR-GNN with other SBR models.

\subsubsection{SBR Modeling}
In this section, we use the SBR model to model interactive information in conversation sequences and learn user behavior preferences. The SBR model here can be replaced arbitrarily. Given that SR-GNN \cite{wu2019session} stands as one of the classic models in SBR, and the state-of-the-art (STOA) models in SBR predominantly rely on GNN, this model holds significant importance. Therefore, we primarily select it as the prototype SBR model within the framework for the experimental segment. Specifically, SR-GNN constructs session data into a session graph, where each node in the graph represents a unique item in the session. It utilizes GGNN to learn node features, then takes the last clicked item in the session as the local embedding of the session. It aggregates all node information and utilizes a soft attention mechanism to represent global preferences.
\begin{equation}
    H_t^l, H_t^g = \textbf{\text{SBR}\textit{-}\text{Model}}(\mathcal{I}_t),
\end{equation}
where $\mathcal{I}_t \subseteq \mathcal{I}$ represents the set of items interacted with in session at time $t$. $H_t^l$, and $H_t^g$ represent the local embedding and global embedding of session $t$ respectively.

\subsubsection{Representation Alignment and Fusion}
The SBR model models the interaction information in a session, while the LLM uses its knowledge to infer the semantic content corresponding to the session. Although both have the same goal, they are not in a unified embedding space. 
Here we draw on the approach of DirectAU \cite{wang2022towards} to unify the SBR and LLM representations into a common space to achieve alignment and unification of the representations in different viewpoints.

\begin{equation}
    \Tilde{h_{\text{infer}}^\text{p}} = \mathbf{T} h_{\text{infer}}^\text{p},
\end{equation}
\begin{equation}
    \mathcal{L}_{a} = \underset{(\text{infer},\text{t}) \sim sess}{\mathbb{E}}||\Tilde{h_{\text{infer}}^\text{p}} - h_\text{t}^\text{p} ||^2, \label{Align}
\end{equation}
\begin{equation}
    \mathcal{L}_{u} = {\log e^{-2 {||\Tilde{h_{\text{infer}}^\text{p}} - \Tilde{h_{\text{infer}^{'}}^\text{p}}||}^{2}}/2} + {\log e^{-2 {||h_{t}^\text{p} - h_{t^{'}}^\text{p}||}^{2}}/2} , \label{Uniform}
\end{equation}
where the matrix $\mathbf{T} \in \mathbb{R}^{d \times d_{\text{text}}}$ transforms the embedding vectors of the text modal into the latent space $\mathbb{R}^{d}$, $\mathcal{L}_{a}$ denotes alignment loss function and $\mathcal{L}_{u}$ denotes uniformity loss function. 
For each view (long-term, short-term), we separately compute the alignment loss between the inference representation and session representation under that view and the uniform loss within each inference representation and each session representation. 

We then fuse session representations from different views and modalities into a final session representation through a soft-attention mechanism.
\begin{equation}
    \alpha_{\text{infer}}^{\text{lt}} = \textbf{Q}_1^\mathrm{T}(\text{sigmoid}(H_\text{t}^{g} + H_{\text{infer}}^{\text{lt}})),
\end{equation}
\begin{equation}
    \alpha_{\text{infer}}^{\text{st}} = \textbf{Q}_2^\mathrm{T}(\text{sigmoid}(H_\text{t}^{l} + H_{\text{infer}}^{\text{st}})),
\end{equation}
\begin{equation}
    H_{\text{sess}} = \textbf{W} [ H_\text{t}^{l} * \alpha_{\text{infer}}^{\text{st}} ; H_\text{t}^{g} * \alpha_{\text{infer}}^{\text{lt}}], \label{conbine}
\end{equation}
where parameters $\textbf{Q}_1,\textbf{Q}_2 \in \mathbb{R}^d$ and $\textbf{W} \in \mathbb{R}^{d \times 2d}$ transforms the concatenated embedding vectors into a latent space $ \mathbb{R}^{d}$. $H_\text{t}^{l}$ is the local preference representation obtained in the SBR model, where the local preference embedding is simply defined as the last clicked item. $H_\text{t}^{g}$ is the global embedding obtained by the SBR model, which is obtained by the soft attention mechanism. For details, please see SR-GNN \cite{wu2019session}. Additionally, $H_{\text{infer}}^{\text{st}}$ and $H_{\text{infer}}^{\text{lt}}$ represent the short-term and long-term view text embeddings of LLM inference, respectively.
Finally, we concatenate the weighted local and global representations into the final session representation $H_{\text{sess}}$.

\subsubsection{Prediction and Optimization}
By taking the item of the session representation and the item representation, scores for each candidate item are obtained. Then, the softmax function is applied to obtain the model's predicted values $Y$.
\begin{equation}
    \hat{y_i} = \text{softmax}(h_{\text{sess}}^\mathrm{T} v_{i}),
\end{equation}
where $\hat{y_i}$ represents the probability that each item in the itemset becomes the next item in the current session. The loss function for SBR tasks is defined as the cross-entropy between the predicted values and the ground truth, as shown below:
\begin{equation}
    \mathcal{L}_{r} = - \sum_{i=1}^n y_i log(\hat{y_i})+(1-y_i)log(1-\hat{y_i}),
\end{equation}
where $y$ is the one-hot encoding vector of the ground truth item.

Ultimately, the joint learning loss function is composed of both the recommendation loss function and the auxiliary task (alignment and uniformity) loss function.
\begin{equation}
    \mathcal{L} = \mathcal{L}_{r} + \tau \left( \mathcal{L}_{a} + \mathcal{L}_{u} \right), \label{eq:opt}
\end{equation}
where $\tau$ controls the proportion of auxiliary tasks.

\subsection{Discussion}

\textbf{Ease-of-Use:} One key aspect that makes LLM4SBR easy to use is its modular design. The separation of Intent Inference and Representation Enhancement stages allows developers to integrate or modify individual components without necessarily disrupting the entire system. Moreover, LLM4SBR retains the data interface of the traditional SBR model, which means that you only need to prepare the data required for LLM inference to seamlessly transition to the training phase without making large-scale changes to the existing workflow. This makes the migration from the old system to the new model smoother and more efficient.

\textbf{Plug-and-Play Compatibility:}
The LLM4SBR framework utilizes SBR models to process interaction data, ensuring that advances in session modeling can be easily adopted without the need for radical modifications. This compatibility encourages researchers to experiment with different SBR models as ``plug-ins,'' choosing the most appropriate model based on their specific application requirements or performance benchmarks. 
Additionally, the LLM in the LLM4SBR framework can also be updated and does not require dedicated pre-training and fine-tuning. As research in the LLMs continues to advance, the LLMs in the framework can be replaced by newer and more powerful models. This feature enables it to adapt to the ever-changing research environment and technological advances.

\section{Experiments}
In this section, we design a series of experiments to answer the following four questions:
\begin{itemize}
    \item \textbf{RQ1:} Can the LLM4SBR framework improve the performance of the SBR model?
    \item \textbf{RQ2:} Is each component in the LLM4SBR framework necessary and what is its impact on the overall performance?
    \item \textbf{RQ3:} How do hyperparameter settings affect performance in LLM4SBR?
    \item \textbf{RQ4:} Is the LLMSBR framework lightweight?
\end{itemize}

\label{sec::exp}
\subsection{Experimental Settings}
\subsubsection{Datasets}
\begin{table}[t]
    \centering
    \caption{Statistics of the utilized datasets.}
    \resizebox{0.5\textwidth}{!}{
    \begin{tabular}{c|ccccc} \hline
    \textbf{Datasets}   & \textbf{Train}   & \textbf{Test}   & \textbf{Clicks}  & \textbf{Items} & \textbf{Avg.len.}   \\ \hline
    Beauty & 158,139 & 18,000 & 198,502 & 12,101 & 8.66  \\ 
    Ml-1M & 47,808  & 5,313  & 987,610  & 3,416 & 17.59  \\ \hline
    \end{tabular}
    }
    \label{table 1}
\end{table}

We initially hoped to use datasets commonly used in SBR, such as Diginetica, Nowplaying, and RetailRocket, etc., to validate performance as they are more representative. Unfortunately, none of these datasets provide both interaction ID sequences and item name information. Consequently, we selected the classic datasets Beauty and MovieLens-1M (Ml-1M) in the Sequential Recommendation, and adapted them to a session format.

The details of these two datasets are shown in Table \ref{table 1}. For both datasets, we adhere \cite{wu2019session, li2017neural} to removing sessions with a length of $1$ and items that appear fewer than $5$ times across all sessions.
\begin{itemize}[leftmargin=*]
    \item \textbf{Beauty} \footnote{https://jmcauley.ucsd.edu/data/amazon/links.html} dataset comprises evaluations and ratings from users on various beauty items. We treat all ratings sequences from a single user as a session sequence. We enhance the dataset using the commonly employed sequence segmentation method \cite{li2017neural,liu2018stamp,wu2019session} in SBR.
    For instance, consider an original session $s=[i_{t,1}, i_{t,2}, \cdots, i_{t,n}]$. After segmentation by sequence, we obtain $([i_{t,1}],i_{t,2}), ([i_{t,1}, i_{t,2}],i_{t,3})$, $\cdots$ ,$([i_{t,1},i_{t,2},\cdots,i_{t,n-1}],i_{t,n})$.

    \item \textbf{Ml-1M} \footnote{https://grouplens.org/datasets/movielens/} dataset consists of over $1$ million ratings from more than $6,000$ users on over $4,000$ movies. Considering that our research problem is SBR, we have performed special processing on the sequence dataset \textbf{by using 10-minute intervals as segmentation points to divide the user sequence into multiple session sequences.}
\end{itemize}

\subsubsection{Evaluation metrics}
In terms of the evaluation indicators used in the experiment, We chose the most commonly used ones in SBR tasks: Precision (P) $@K$, Mean reciprocal rank (MRR) $@K$ and Normalized Discounted cumulative gain (NDCG) $@K$.
After referring to the classic work \cite{wu2019session, wang2020global, xia2021self, huang2021graph} \footnote{The calculation method of $P@20$ and $MRR@20$ refer to the \cite{wu2019session}, and $NDCG@K$ refers to \cite{huang2021graph}.} in recent years, we set the length of the candidate set $@K$ to $5$, $10$, and $20$, which is the most meaningful for comparison.

\subsubsection{Implementation details and Hyper-parameter settings}
All experiments were conducted on NVIDIA A100 GPUs. 
In the aspect of selecting large models, we have chosen the Qwen-7B-Chat \footnote{https://github.com/QwenLM/Qwen} model as the inference model after careful consideration of LLM's inference capability, adaptability to both Chinese and English languages and model parameter count.
For fairness in performance comparison, the optimizer used throughout the experiments was unified as Adam with a learning rate of $0.001$, decayed by $0.1$ every three epochs, and an $L2$ penalty set to $10^{-5}$. For the SBR model involved in the experiments, the batch size is $100$ and the dimension size is $100$. $\tau$ is set to 0.1. We initially set the hyperparameter $r$ in the intent localization module to $5$, and subsequent hyperparameter experiments \ref{sec:Hyper} will discuss the optimal value.
We followed the optimal parameter settings as published in their paper for the remaining parameters.

\subsection{Performance Experiment and Analysis (RQ1)}
In this section, we mainly compare the performance of the SBR model and the corresponding SBR model applying the LLM framework under different Top-$K$. 
\subsubsection{Backbones}
\label{sec::experiments::baselines}
To validate the effectiveness of the framework, we carefully selected six classic models in the field of Session-based Recommendation/Sequential Recommendation, with the SBR model as its core component. 
Note that here we have not chosen to compare performance with a model that incorporates LLM. This is because there is very little work on SBR incorporating LLM, and no open-source code exists. Second, the architecture of our work is different from other work, focusing on enhancements to the traditional SBR model rather than from the view of replacing SBR with LLM.
Among them, the GRU4Rec and SASRec models are particularly good at mining sequences' temporal dynamics and patterns. In contrast, SRGNN and TAGNN focus more on revealing the transfer or transformation relationships between items. The GCE-GNN and DHCN models, based on effectively capturing item transformation relationships, further incorporate the concept of multi-graph structure and introduce the use of inter-session information to enhance the model's ability to understand complex interaction patterns. In the experimental section, we assess the performance of each model both before and after enhancement with the framework and systematically compare and analyze the performance disparities among the enhanced models. 

The introduction of the SBR models is as follows:
\begin{itemize}[leftmargin=*]
    \item \textbf{GRU4Rec} \cite{hidasi2015session} utilizes GRU to learn dependencies between sequences to predict the next likely item of interest.
    \item \textbf{SASRec} \cite{kang2018self} uses a self-attention mechanism to solve the long-term dependency problem existing in traditional sequence models.
    \item \textbf{SR-GNN} \cite{wu2019session} is the first model to construct data into session graphs, utilizing GGNN to capture complex transition relationships among items.
    \item \textbf{TAGNN} \cite{yu2020tagnn} adds a target-sensitive attention mechanism based on SR-GNN.
    \item \textbf{GCE-GNN} \cite{wang2020global} constructs session graphs and global graphs respectively, and learns relevant information from the item level and session level.
    \item \textbf{$S^2$-DHCN} \cite{xia2021self} uses hypergraph convolution to learn high-order relationships in item sequences, and uses self-supervised learning to alleviate the data sparse problem of hypergraphs.
\end{itemize}

\begin{table*}[t]
    \centering
    \caption{Performance comparison experimental results (\%).}
    \label{table 2}
    \resizebox{1.0\textwidth}{!}{
    \begin{tabular}{ccccccc|cccccc}
    \toprule
     Dataset& \multicolumn{6}{c}{Beauty} &\multicolumn{6}{c}{Ml-1M}  \\
    \midrule
    Model & $P@5$& $P@10$& $P@20$ &$MRR@5$ &$MRR@10$ &$MRR@20$& $P@5$& $P@10$& $P@20$ &$MRR@5$ &$MRR@10$ &$MRR@20$\\
    \midrule
    GRU4Rec &4.99 &8.18 &12.71 &2.71 &3.06& 3.23&5.48 &9.52 &14.72 &2.98 &3.40& 3.57\\
    LLM4SBR(GRU4Rec) &7.50 &11.68 &16.74 &4.51 &4.89& 4.99&6.66&10.91 &17.20 &3.97 &4.38& 4.52\\
    GRU4Rec Improv. 
    &50.30\%&42.96\%&31.68\%&66.42\%&60.32\%&54.60\% &21.53\%&14.70\%&16.88\%&33.22\%&28.71\%&26.72\% \\
    \midrule
    SASRec &3.93 &6.40 &10.11 &2.09 &2.37& 2.54&2.07 &3.67 &6.47 &1.07 &1.27& 1.42\\
    LLM4SBR(SASRec) &5.29 &8.50 &13.20 &2.98 &3.33& 3.49&3.44&6.06 &10.05 &1.72 &2.02& 2.21\\
    SASRec Improv. 
    &34.6\%&32.81\%&30.56\%&42.58\%&40.51\%&37.94\% &66.18\%& 65.12\%&55.33\%&60.74\%&59.06\%&55.63\% \\
    \midrule
    SR-GNN &6.23 &10.06 &15.06 &3.45 &3.84& 3.99&4.42 &7.53 &12.27 &2.44 &2.80& 2.98\\
    LLM4SBR(SR-GNN) &7.78 &11.73 &16.98 &\textbf{4.71} &\textbf{5.03}& 5.12&8.13&11.80 &18.35 &4.85 &5.16& 5.27\\
    SR-GNN Improv. 
    &24.87\%&16.60\%&12.74\%&36.52\%&30.98\%&28.32\% &83.93\%&56.70\%&49.55\%&98.77\%&84.28\%&76.84\% \\
    \midrule
    TAGNN &6.12 &10.06 &15.23 &3.10 &3.63& 3.97&3.60 &6.19 &10.28 &1.77 &2.15& 2.23\\
    LLM4SBR(TAGNN) &\textbf{7.79} &11.79 &16.76 &4.39 &4.78& 5.05&7.47 &12.33 &18.60 &4.03 &4.79& 4.87\\
    TAGNN Improv. 
    &27.28\%&17.19\%&10.04\%&41.61\%&31.68\%&27.20\% &107.5\%&99.19\%&80.93\%&127.68\%&122.79\%&118.38\% \\
    \midrule
    GCE-GNN &6.39 &8.93 &12.38 &\underline{3.97} &\underline{4.30}& \underline{4.54}&5.16 &6.85 &9.67 &3.18 &3.41& 3.60\\
    LLM4SBR(GCE-GNN) &7.75 &\textbf{12.48} &\textbf{18.08} &3.91 &4.41& 4.80&7.10 &13.44 &22.10 &3.14 &3.63& 4.21\\
    GCE-GNN Improv. 
    &21.28\%&39.75\%&46.04\%&-1.51\%&2.56\%&5.73\% 
    &37.59\%&96.20\%&128.54\%&-1.25\%&6.45\%&16.94\%\\
    \midrule
    $S^2$-DHCN &\underline{7.14} &\underline{11.97} &\underline{17.54} &2.97 &3.61 &  3.99&\underline{8.35} &\underline{14.55} &\underline{\textbf{23.38}} &\underline{3.66} &\underline{4.51} & \underline{5.09}\\
    LLM4SBR($S^2$-DHCN)&7.77 &11.85 &17.48 &4.26 &4.79& \textbf{5.15}&\textbf{9.54} &\textbf{15.31} &22.67 &\textbf{5.13} &\textbf{5.91}& \textbf{6.40}\\
    $S^2$-DHCN Improv. 
    &8.82\%&-1.00\%&-0.34\%&43.43\%&32.68\%&29.07\% 
    &14.25\%&5.22\%&-3.03\%&40.16\%&31.04\%&25.73\%\\
    \bottomrule
    \end{tabular}
    }
    \begin{tablenotes}
       \footnotesize
       \item[1] * We highlight the best performance values for each metric in bold and underscore the best values within the backbones. 
       \item[2] * The calculation formulas for $P@K$, $HR@K$ and $Recall@K$ are the same in SBR. 
     \end{tablenotes}
\end{table*}

The comparison results of the overall performance experiments are shown in Table \ref{table 2}. 
We record the performance with K set to {5, 10, 20}. It is worth noting that smaller $K$ values are more significant in the evaluation system of RS. From the results displayed in Table \ref{table 2}, we draw the following observations:
\begin{itemize}[leftmargin=*]
    \item \textbf{LLM4SBR significantly improves backbone performance.} In the models enhanced through the LLM framework, both sequence and graph-structured models show significant performance improvements. For example, the $P@5$ of SASRec and SR-GNN on the Ml-1M dataset have increased by 66.18\% and 83.93\%, respectively. This confirms that the text representations derived from LLM inference contain rich and valuable information, which can greatly help the SBR model understand the potential intention of the conversation data. 
    \item \textbf{LLM4SBR has a greater improvement for smaller $K$ values.} Almost all backbone models achieve larger improvements at smaller $K$ values when combined with the framework. For example, LLM4SBR (TAGNN) improved the $P@5$ index of the two data sets by 27.28\% and 107.5\% respectively. We believe this is due to the semantic enhancement achieved by LLM4SBR during the intent localization stage, where it utilizes $r$ similar semantic items. Consequently, it results in more accurate predictions for the top few items in the predicted candidate set. We also observe slight decreases in performance for $S^2$-DHCN and GCE-GNN on a few metrics ($P@20$ and $MRR@20$) after integrating with the framework. We posit that when the original SBR model already effectively models the data, enhancing the inference information through the intent localization module may introduce noise. Compared to the improvement magnitude, the decrease is very slight. Moreover, since noise issues can be effectively controlled by adjusting the hyperparameter $r$ in the intent localization module, the negative impact can be almost negligible.

    \item \textbf{The LLM4SBR framework can effectively alleviate modeling defects caused by scene mismatch and data sparseness and fully activate the inherent potential of the model.} Due to the short sequence characteristics of session data, although SASRec is good at capturing long-term dependencies, this becomes a limitation in short sequence scenarios, resulting in poor performance. However, after being integrated into the LLM4SBR framework, its shortcomings in short-sequence processing have been significantly compensated, and its performance has been greatly improved. This achievement not only reflects the efficiency of framework optimization and collaborative model capabilities but also proves that framework integration can effectively enhance the model potential in specific scenarios. GCE-GNN captures effective information at both the item and session levels by constructing global graphs and session graphs simultaneously, due to the model's complex computations, in scenarios with limited data volume, it becomes challenging for this model to learn effective session representations. LLM4SBR (GCE-GNN) showed the greatest improvement, especially on the Ml-1M dataset, $P@5$, $P@10$, and $P@20$ increased by 37.59\%, 96.2\%, and 128.54\% respectively. We attribute this to the effective text information obtained from LLM inference, which compensates for the information scarcity in GCE-GNN's session modeling, allowing it to achieve better performance. 
    
\end{itemize}

In conclusion, the effectiveness of the LLM4SBR framework is undeniable. As a plug-and-play framework, it significantly enhances the prediction accuracy of traditional SBR models. 

\subsubsection{Comparison with SBR Models Combining LLMs}

\begin{figure}
    \centering
    \begin{subfigure}{0.8\textwidth}
    \centering
    \includegraphics[width=\textwidth]{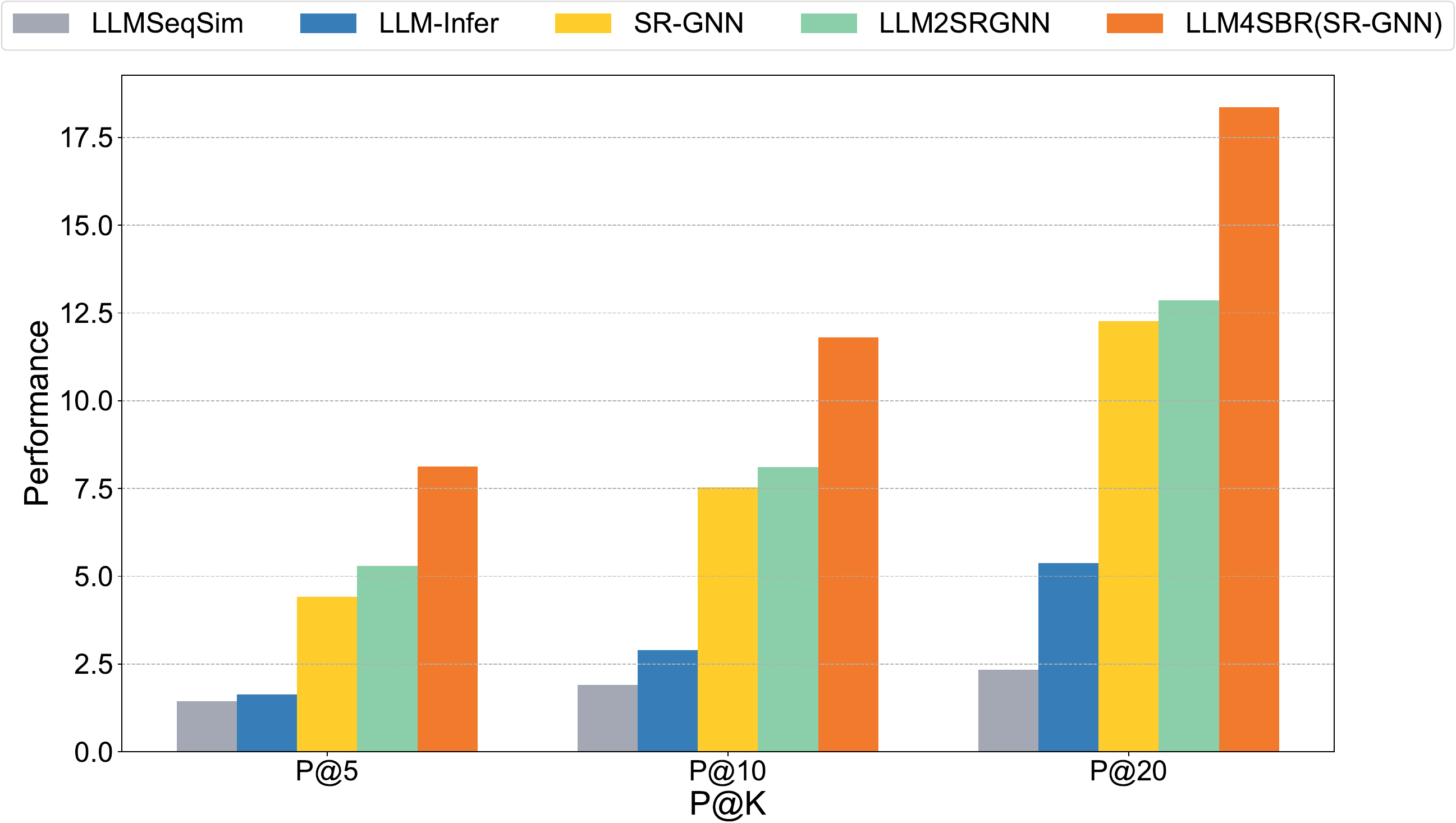}
    \label{fig:LLM_P}
  \end{subfigure}
  \begin{subfigure}{0.8\textwidth}
    \centering
    \includegraphics[width=\textwidth]{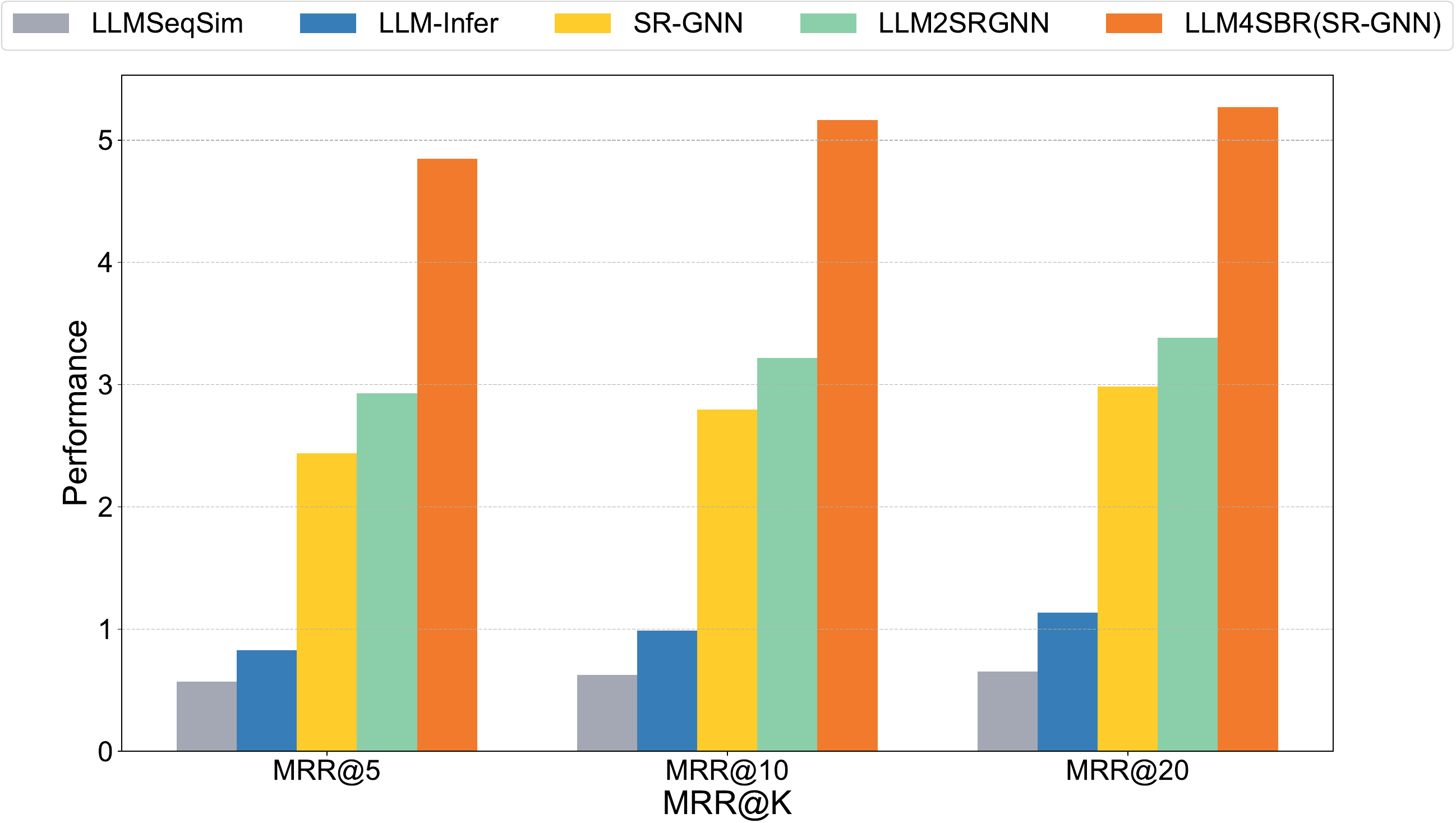}
    \label{fig:LLM_M}
  \end{subfigure}
  \caption{Performance Comparison with SBR Models Combining LLMs.}
    \label{fig:LLM}
\end{figure}
In this subsection, we further compare the performance of the LLM4SBR framework with other frameworks that combined with LLM to explore the enhancement capability of LLM in LLM4SBR for SBR models. Note that here we use the same backbone, SRGNN. the specific model is described below:
\begin{itemize}
    \item \textbf{LLM-Infer} directly predicts the next click using LLM inference of long-term and short-term interests after intent localization.
    \item \textbf{LLMSeqSim}\cite{harte2023leveraging} retrieves semantically rich embeddings for each item in the session from existing LLMs and computes aggregated session embeddings to recommend products with similar embeddings.
    \item \textbf{LLM2SRGNN} \cite{boz2024improving}\footnote{We used the LLM2Sequential framework but replaced the SR model backbone with SR-GNN to ensure the fairness of the experiments.} initializes the SBR model with the item embeddings obtained from LLM.
\end{itemize}

In Figure \ref{fig:LLM}, we show the original SR-GNN model in yellow. It can be clearly seen from the figure that LLM-Infer represented in gray and LLMSeqSim represented in blue are significantly lower in performance than SR-GNN, which effectively models collaborative information. This observation demonstrates that although LLMs perform well on many NLP tasks, they are not suitable for direct application in recommendation tasks. At the same time, the other two frameworks (LLM2SRGNN and LLM4SBR) that use LLM to enhance the collaborative model in the figure have achieved significant performance improvements compared to the backbone (SR-GNN). This experimental result confirms that combining LLM with the collaborative model is a wise direction. LLM can provide rich semantic modality information for the collaborative model that is limited by data sparsity. It is particularly noteworthy that LLM4SBR achieved the best results among all models, and achieved a significant performance improvement compared to the second-best SR-GNN, which proves the rationality and effectiveness of our framework design. We conclude that it is not enough to simply introduce knowledge from LLM, but also to effectively integrate information from different modalities and improve the effective utilization of information to achieve more accurate recommendation results.

\subsection{Ablation Study (RQ2)}
\subsubsection{The impact of LLMSBR framework integration on performance}

\begin{table}[]
    \caption{The ablation results of LLM and SBR models in the LLM4SBR framework.}
    \resizebox{0.9\textwidth}{!}{
    \centering
    \begin{tabular}{c|cccc|cccc}
        \hline
     Dataset& \multicolumn{4}{c}{Beauty} & \multicolumn{4}{c}{Ml-1M} \\
    \hline
    Model & w/o SBR& w/o LLM& w/o AU &LLM4SBR& w/o SBR& w/o LLM&w/o AU & LLM4SBR \\
    \hline
    $P@5$  & 1.92&6.23 &7.12 &\textbf{7.78} &1.64& 4.42 &6.31&\textbf{8.13} \\
    $P@10$  & 3.44&10.06 &10.75&\textbf{11.73}& 2.90&7.53 &10.07&\textbf{11.80}  \\
    $P@20$ & 5.76&15.06 & 15.74&\textbf{16.98}&5.38& 12.27&15.11&\textbf{18.35} \\
    \hline  
    $MRR@5$  & 0.95&3.45&4.39 &\textbf{4.71} & 0.83&2.44 &3.85&\textbf{4.85} \\
    $MRR@10$  & 1.14&3.84&4.69 &\textbf{5.03}& 0.99&2.80 & 4.20&\textbf{5.16}  \\
    $MRR@20$ & 1.27&3.99 & 4.79&\textbf{5.12}&1.13& 2.98&4.34&\textbf{5.27} \\
    \hline
    $NDCG@5$  & 1.21&4.27 &5.21&\textbf{5.64} & 1.04&3.02 &4.54&\textbf{5.89} \\
    $NDCG@10$  & 1.69&5.38 &6.22&\textbf{6.67}& 1.44&3.98 &5.60&\textbf{6.90}  \\
    $NDCG@20$ & 2.23&6.39 & 7.16&\textbf{7.64}&2.03& 4.97&6.65&\textbf{8.10} \\
    \hline
    \end{tabular}
    }
    \label{tab:3}
\end{table}
To test the integration effect of the LLM4SBR framework, we design three variants as follows: 
\begin{itemize}
    \item \textbf{LLM4SBR w/o SBR} - using the long-term and short-term interest results of LLM inference after intention localization for direct prediction.
    \item \textbf{LLM4SBR w/o LLM} - using the SBR model (here we use SR-GNN) for direct prediction.
    \item \textbf{LLM4SBR w/o AU} - removing the representation alignment operation, and directly fuses the LLM inference representation with the long and short-term session representation constructed by SBR.
\end{itemize}
We added $NDCG@K$ as an evaluation criterion on top of the $P@K$ and $MRR@K$ evaluation metrics. Then we compared the performance of these three variants with the full version of LLM4SBR on the Beauty and Ml-1M datasets. The experimental results are shown in Table \ref{tab:3}, where we marked the best-performing model in bold.

By observing and analyzing the results in Table \ref{tab:3}, we draw the following conclusions:
\begin{itemize}
    \item \textbf{LLM inference results are ineffective when directly used for recommendation tasks.} In the experimental results shown in Table \ref{tab:3}, the LLM4SBR w/o SBR variant performed poorly, ranking last across all evaluation metrics. This finding is consistent with previous research \cite{bao2023tallrec, gao2023chat}, which suggests that there is a significant difference between pre-trained LLMs and the tasks required for the RS, leading to their limited effectiveness when directly applied to SBR scenarios. RSs require models to understand user historical behavior patterns, capture real-time changes in user interests, and effectively match associations within a high-dimensional item space. In contrast, LLMs primarily focus on coherence and meaning expression within language structures, lacking the in-depth understanding of domain-specific knowledge and context required for the recommendation field. Therefore, the core of this study is the innovative integration strategy, which aims to combine the powerful language understanding of LLM with the targetedness of traditional recommendation algorithms to build a more efficient and intelligent RS.

    \item \textbf{The LLM4SBR framework effectively combines the advantages of SBR and LLM.} The SBR model using the LLM4SBR framework has the best performance in Table \ref{tab:3}, in which the $NDCG@K$ metrics are significantly enhanced. This improvement demonstrates the advantages of LLM4SBR in accurately capturing user interests and proves that the LLM4SBR framework can skillfully integrate the expertise of LLM and SBR to achieve high efficiency in collaborative modeling. The second best performance of all metrics is the LLM4SBR w/o AU, which firstly proves the validity of the operation of split-view representation alignment in the framework, which is more scientifically sound to align and uniform the representations of different modalities in separate views before fusing them, than to directly fuse the different modal representations, and secondly, this variant proves once again that the results of the LLM inference can indeed significantly enhance the performance of the SBR, even without the representation alignment operation, it also performs better than the modeling approach using unimodal information. Through the LLM4SBR framework, the SBR model can not only deeply understand the text context and extract more detailed and comprehensive user preferences, but also effectively handle sequence dependencies in conversational recommendation scenarios, thereby generating more personalized recommendation results.
\end{itemize}

\subsubsection{The influence of LLM inference results from different views}
\begin{figure}
    \centering
    \begin{subfigure}{0.48\textwidth}
    \centering
    \includegraphics[width=\textwidth]{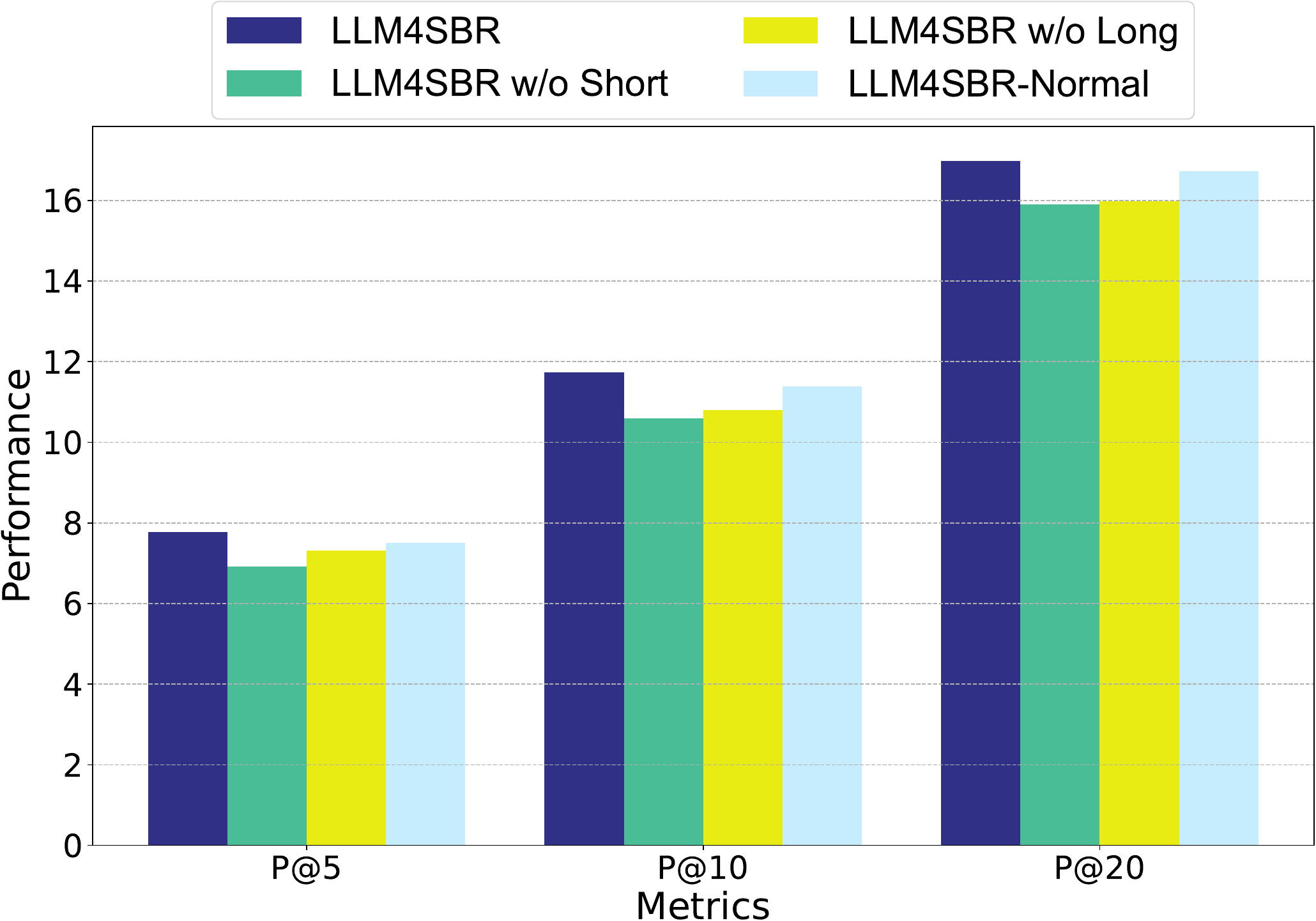}
    \caption{$P@K$ on Beauty.}
    \label{fig:sub1}
  \end{subfigure}
  \begin{subfigure}{0.48\textwidth}
    \centering
    \includegraphics[width=\textwidth]{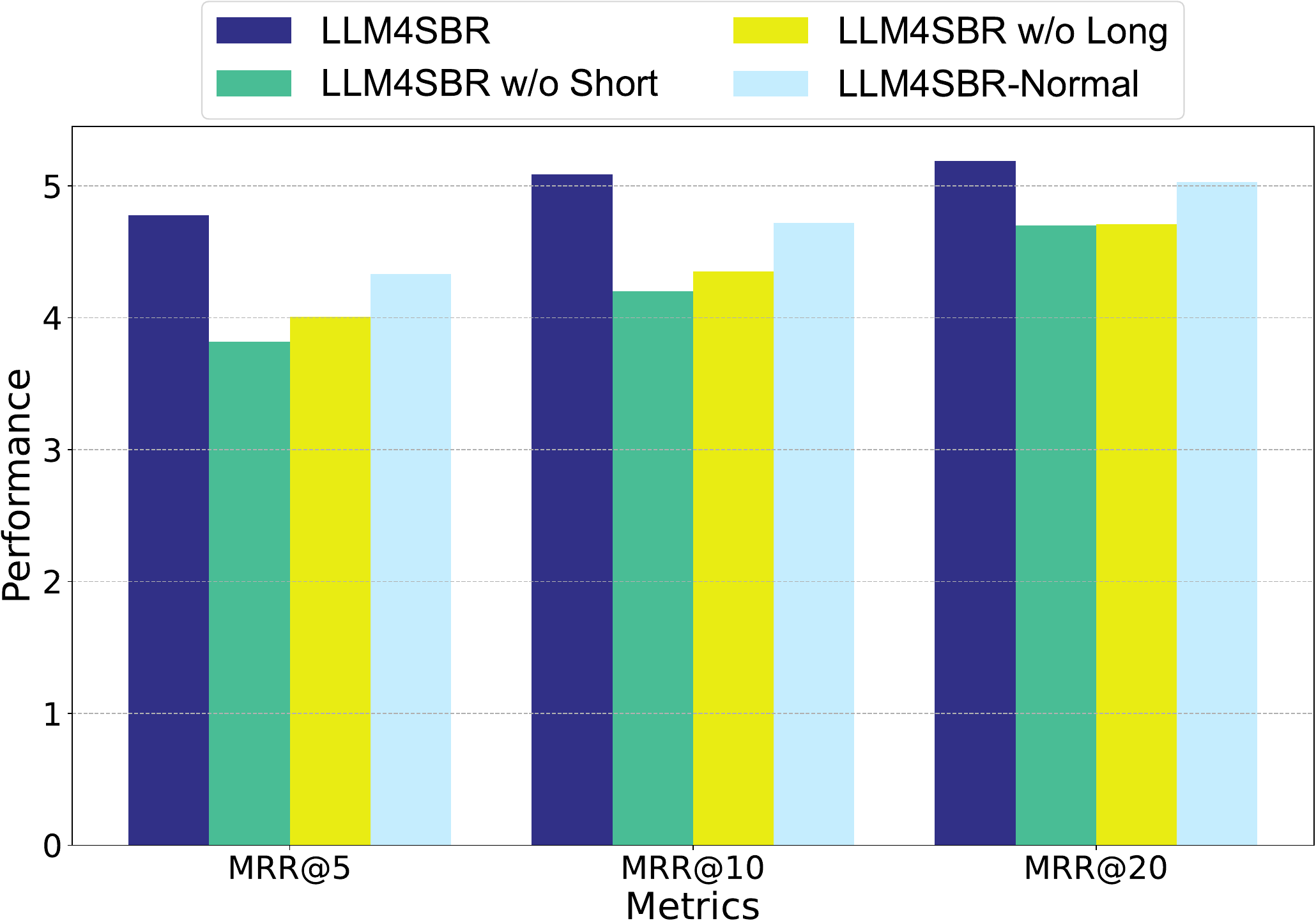}
    \caption{$MRR@K$ on Beauty.}
    \label{fig:sub2}
  \end{subfigure}
  \begin{subfigure}{0.48\textwidth}
    \centering
    \includegraphics[width=\textwidth]{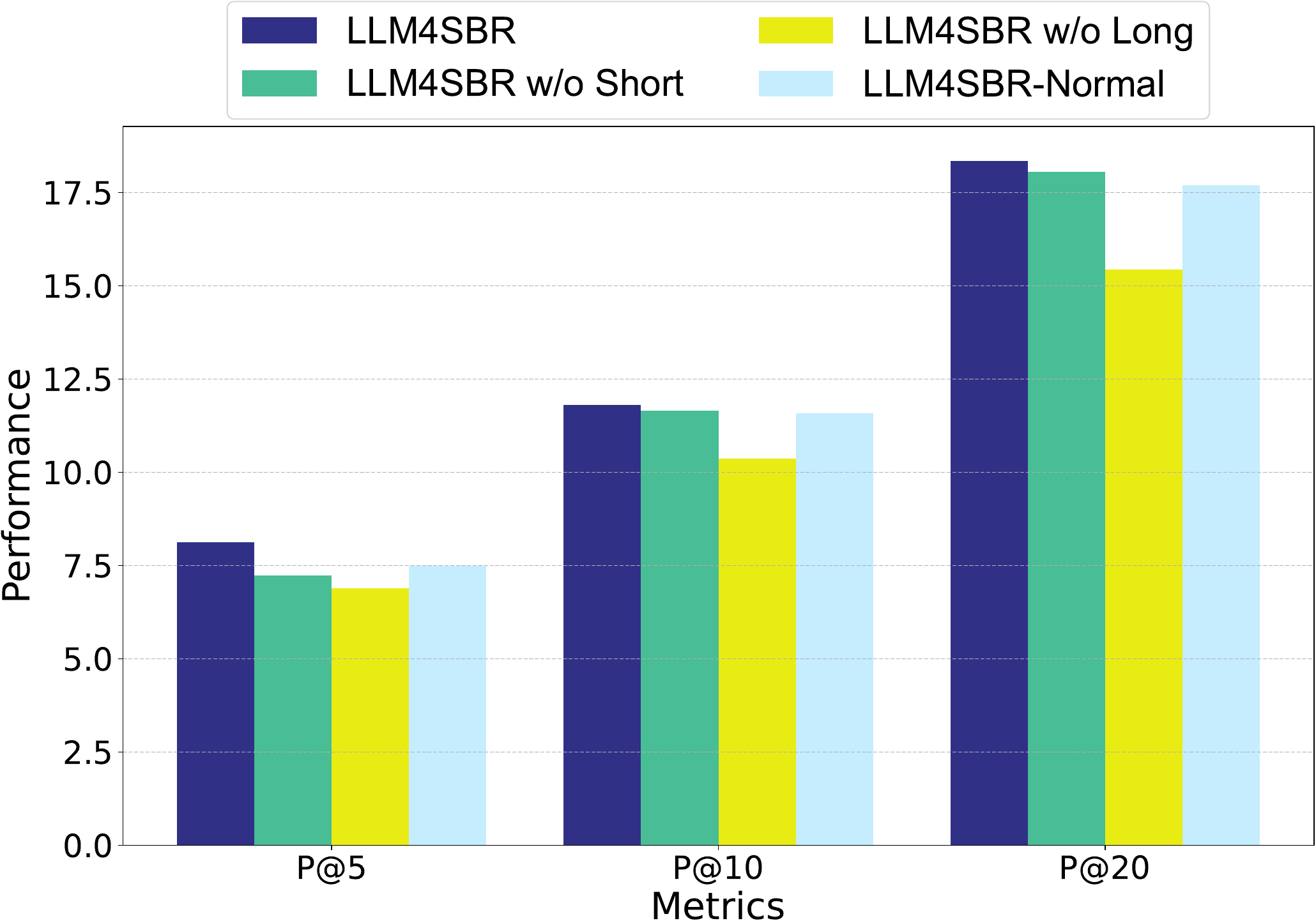}
    \caption{$P@K$ on Ml-1M.}
    \label{fig:sub3}
  \end{subfigure}
  \begin{subfigure}{0.48\textwidth}
    \centering
    \includegraphics[width=\textwidth]{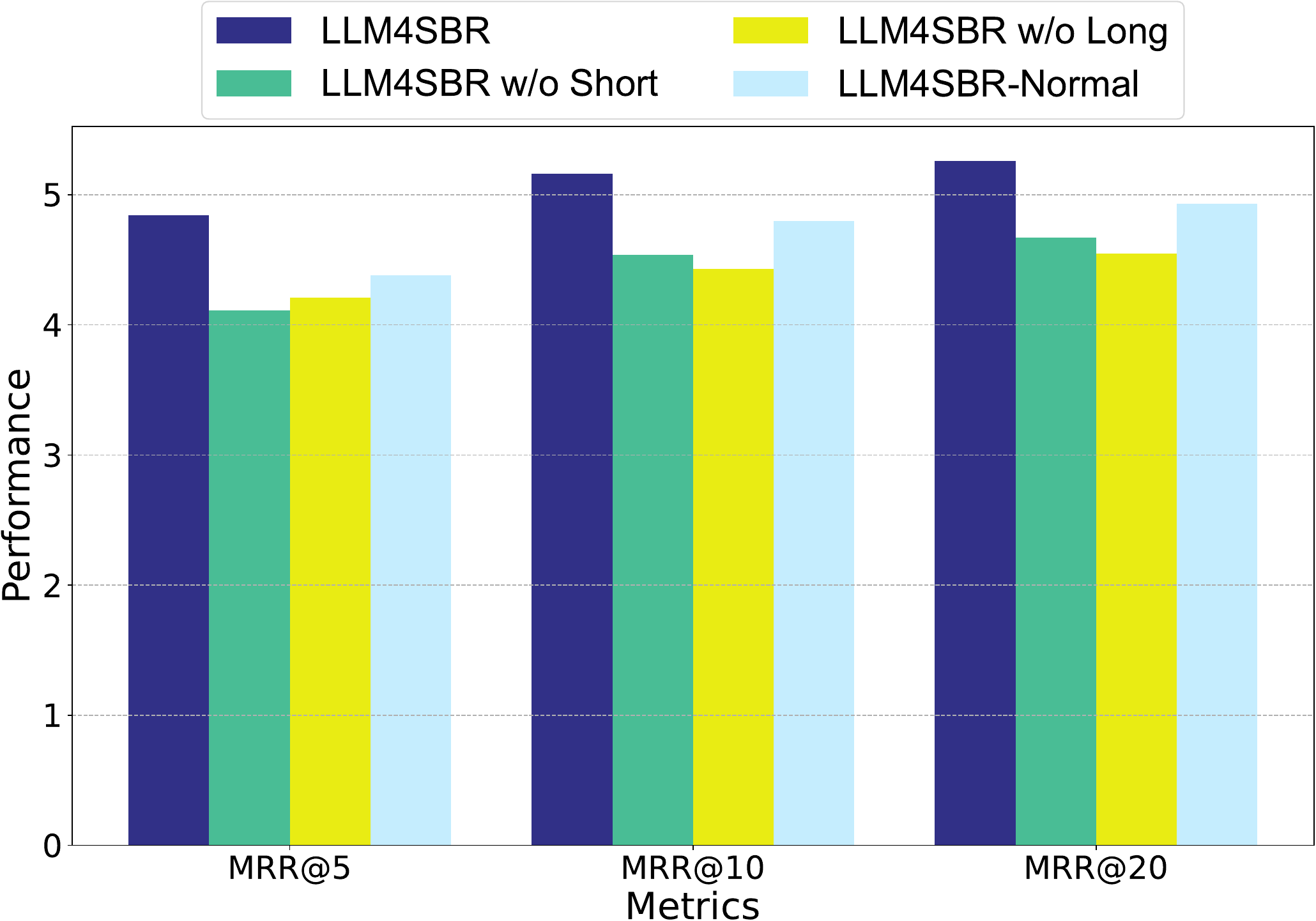}
    \caption{$MRR@K$ on Ml-1M.}
    \label{fig:sub4}
  \end{subfigure}
  \caption{The ablation results from different inference views.}
    \label{fig:4}
\end{figure}

\begin{figure*}[]
  \centering
  \begin{subfigure}{0.3\textwidth}
    \centering
    \includegraphics[width=\textwidth]{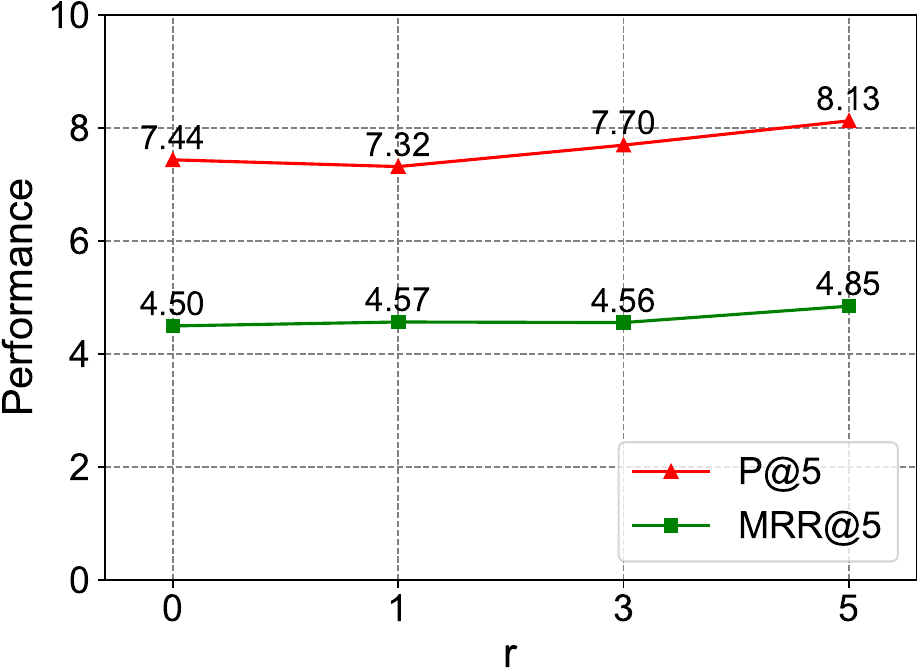}
    \subcaption{The result of $P@5$ and $MRR@5$}
    \label{fig:H_5}
  \end{subfigure}
  \hspace{0.03\textwidth} %
  \begin{subfigure}{0.3\textwidth}
    \centering
    \includegraphics[width=\textwidth]{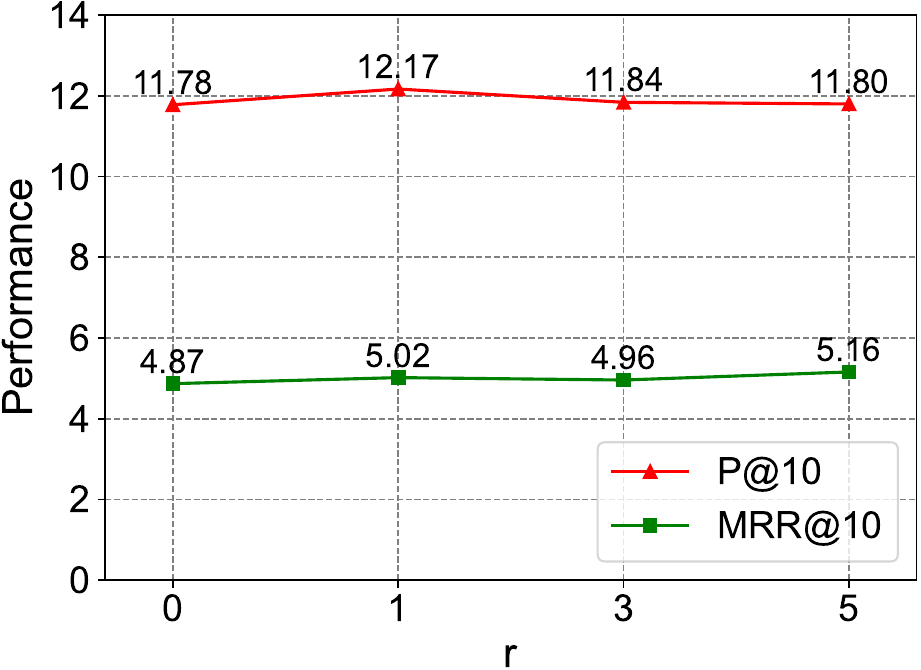}
    \subcaption{The result of $P@10$ and $MRR@10$}
    \label{fig:H_10}
  \end{subfigure}
  \hspace{0.03\textwidth} %
  \begin{subfigure}{0.3\textwidth}
    \centering
    \includegraphics[width=\textwidth]{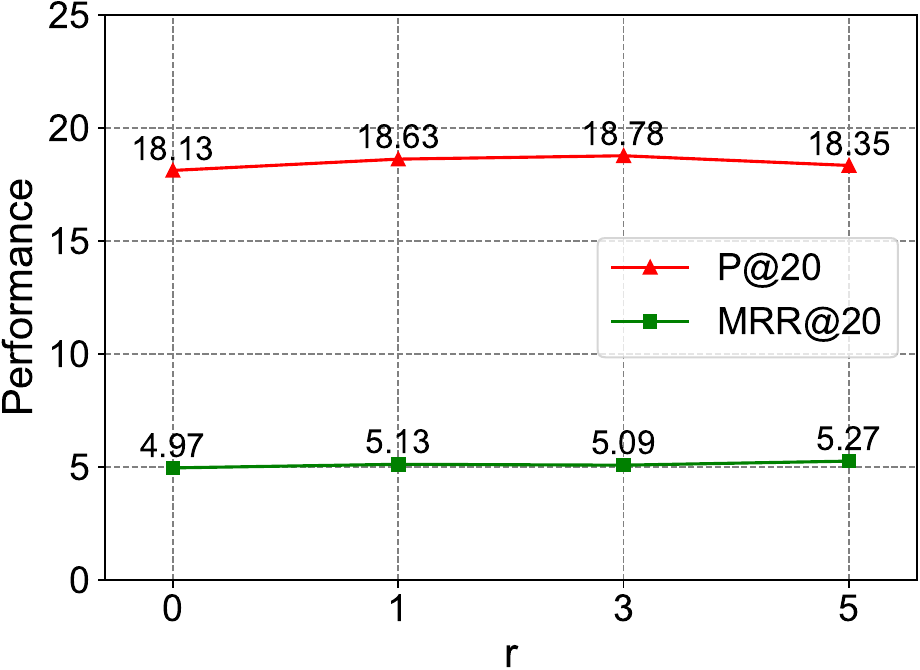}
    \subcaption{The result of $P@20$ and $MRR@20$}
    \label{fig:H_20}
  \end{subfigure}
  
  \caption{Hyperparameter experimental results of different $r$ settings of the intent localization module on Ml-1M.}
  \vspace{0.4cm}
  \label{fig:5}
\end{figure*}

To examine the necessity and relative importance of the long-term and short-term inference views, we designed three variants as follows:
\begin{itemize}
    \item \textbf{LLM4SBR w/o Long} indicates inference without considering the long-term view, retaining only the short-term view.
    \item \textbf{LLM4SBR w/o Short} retains only the long-term view and removes the short-term view during inference.
    \item \textbf{LLM4SBR-Normal} removes the view-limiting qualifier in prompt, and LLM directly infers the overall interest, which is then aligned and uniformed with the comprehensive session embedding learned by the traditional SBR model.
\end{itemize}
We compared the performance of these three variants with the whole performance and visualized the comparison as a bar chart to clearly illustrate the differences between them.

Through observation and analysis of Figure \ref{fig:4}, we summarized the following conclusions:
\begin{itemize}[leftmargin=*]
    \item \textbf{Combining prompts from different views can improve the accuracy of LLM in inferring user interests, both long-term and short-term views are necessary.} Because the whole framework represented by the blue column in the figure shows the best performance on both datasets. For example, on the Beauty, the $P@5$, $P@10$, and $P@20$ of LLM4SBR are 7.78, 11.73, and 16.98, respectively, while the corresponding w/o Long and w/o Short variants show varying degrees of decreases in all three of the above metrics($P@5$, $P@10$, $P@20$), with values of 7.32, 10.80, and 15.98, as well as 6.92, 10.59, and 15.90. This justifies the necessity for multi-view inference, where deleting any view results in a performance loss. 
    In addition, the second best performing model is LLM-Normal, with $P@5$, $P@10$, and $P@20$ metrics of 7.50, 11.38, and 16.72 on the Beauty dataset, which indicates that inference from any single view is inferior to direct comprehensive inference, but combining prompts from different views can improve the accuracy of LLM's inference of user interests and effectively improve the recommendation metrics.

    \item \textbf{The contribution of long-term and short-term view inference varies across the two datasets.} Specifically, on the Beauty dataset, the $P@5$ of LLM w/o Long and LLM w/o Short are 7.32 and 6.92 respectively, and the former performs better than the latter; on the Ml-1M dataset, the $P@5$ of LLM w/o Long and LLM w/o Short are 6.90 and 7.22 respectively, and the former performs worse than the latter. In Beauty, the framework relies more on the information provided by the short-term view, as discarding the inference results of the short-term view would lead to a greater performance drop. Conversely, in Ml-1M, it's the opposite; the framework relies more on the inference results of the long-term view. Through discussion and analysis, we attribute this performance difference to the length of the dataset sessions. Session intent in short sequences is usually relatively stable, and the intent is mainly reflected in the last few clicks. This underscores the increased importance of accurately modeling short-term interests in short-session scenarios. However, as the session length increases, the session intent is influenced by various factors, thereby increasing the importance of long-term dependency relationships within the session. Finally, we believe that simultaneously considering the inference results of multiple views can enhance the stability of the framework's performance, making it adaptable to datasets with varying session lengths.

    \item \textbf{The contribution of view splitting is to enhance the directionality of LLM. } The performance of LLM4SBR-Normal is the second best on both datasets, and the performance degradation is not significant. We analyzed the experimental results and believe that there are two main reasons. First, LLM itself has strong reasoning ability. Without clear guidance, LLM can combine global context information and summarize the user's overall interests on its own. Second, the long-term and short-term view reasoning is mainly used to enhance LLM's fine-grained understanding of user interests, rather than completely changing its reasoning method. By introducing different view qualifiers in the prompts, the framework can more accurately guide LLM to infer users' short-term preferences and long-term interests, supplement the missing long-tail items, and thus give full play to its reasoning ability.

\end{itemize}

In conclusion, within the LLM4SBR framework, each module is indispensable. The LLM4SBR framework successfully integrates the strengths of LLM and SBR models by leveraging LLM for multi-view interest inference and aligning semantic and collaborative information from corresponding views, significantly enhancing recommendation performance.

\subsection{Hyperparameter Experiment and Analysis (RQ3)}
\label{sec:Hyper}
In this section, we discuss the hyperparameter $r$ set within the intent localization module. This hyperparameter is designed to alleviate hallucination and enhance semantics in the preliminary inference results of LLM, using a candidate set of items with similar semantics. 
The reason we chose to conduct this experiment on the Ml-1M dataset is that this dataset does not use the data enhancement method of sequence segmentation, which is more reflective of the actual effect of precise localization and semantic enhancement in real scenarios using items with similar semantics.

The hyperparameter $r$ is configured to control the range of selecting items with similar semantics. The value of $r$ is set to 0, 1, 3, and 5, and we discuss four scenarios accordingly: (1) directly utilizing the inference results of LLM; (2) using the most similar 1 items to alleviate hallucination and enhance semantics; (3) using the most similar 3 items; (4) using the most similar 5 items. 

The experimental results are shown in Figure \ref{fig:5}. Firstly, across all three subfigures, although the optimal hyperparameter values differ for each subfigure, it can be seen that in most cases, the performance is the worst when $r =0$. We believe this is logical and demonstrates the necessity of the intent localization module in the framework. If the results of LLM inference are not processed, hallucinations occurring in some session data may decrease the overall framework performance. Moving on to each of the figures, in Figure \ref{fig:H_5}, the performance enhances as $r$ increases, peaking at $P@5$ and $MRR@5$ when $r = 5$, with $P@5$ and $MRR@5$ at 8.13 and 4.85, respectively. In Figures \ref{fig:H_10} and \ref{fig:H_20}, the peaks are achieved when $r=1$ or $r=3$, respectively. However, $r$ has little effect on those metrics. 
We believe that this is because the $r$-value range set by the framework is small (no more than 5), so the intent Localization module only screens the top few items in text similarity for semantic enhancement. Therefore, when evaluating the Top-$K$ metrics, it has a positive impact on smaller $K$ values but has no obvious impact on larger $K$ values.

In summary, $r$ values from $1$ to $5$ are valid. A smaller value of $r$ can play the efficacy of pinpointing, and a larger value of $r$ can improve the accuracy of the whole recommendation list through semantic enhancement. Depending on the actual situation, choosing different $r$ values can better utilize the effectiveness of the module.

\subsection{Model training space occupancy experiment and Time complexity analysis (RQ4)}

\begin{table}[t]
    \centering
    \caption{The results of GPU usage and training time of model training on the Beauty and Ml-1M datasets.}
    \resizebox{0.8\textwidth}{!}{
    \begin{tabular}{c|c|c|c|c}
    \hline
     \bf Dataset & \multicolumn{2}{c}{ Beauty} & \multicolumn{2}{c}{Ml-1M} \\  \hline
     \bf Model  & SR-GNN & LLM4SBR (SR-GNN) & SR-GNN & LLM4SBR (SR-GNN) \\ \hline
      GPU Usage (MB)  & 1,262 &  1,282 & 1,290 &  1,324 \\
       Single Epoch Time (s)   &  398.08  & 1490.49  & 124.89 & 350.80  \\
       Training Time (s) &  2085.67  & 2987.54 & 932.10 &  1406.79 \\
    \hline
    \end{tabular}
    }
    \label{table 3}
\end{table}

Training LLM-based recommendation models typically require a significant amount of GPU resources and longer training time. To explore the spatial and temporal effectiveness of the LLM4SBR framework, We record the GPU usage during training, the training time for a single epoch, and the training time required to achieve the best performance for SR-GNN and LLM4SBR (SR-GNN).

The results are shown in Table \ref{table 3}. First, we can observe that the GPU occupancy rates of the original SR-GNN and the model using the LLM4SBR framework are very close. We believe that this is because the LLM4SBR framework only requires model training in the second stage, and this process does not require the participation of LLM. The SBR model only needs to parse the LLM inference results into tensor form to participate in training, thus saving a lot of training time and GPU resources. Second, looking at the combined single epoch training time and the time for the model training to reach optimal performance, it can be observed that although the training time of the model using the LLMSBR framework for a single epoch maybe 3-4 times longer than that of the original model, the training time to reach the optimal performance is not much different in comparison with the original model, which illustrates the fact that although the model using the LLM4SBR framework will increase the training time for a single epoch, the model can converge faster and reach optimal performance. 

In addition, we analyze the time complexity of the model. In the first stage of LLM4SBR, each prompt corresponds to a conversation sequence. It is assumed that there are $n$ such prompts, and the length of each prompt is quite short compared to n, which means that the time for a single inference can be regarded as constant time $O(1)$. Given that there are $n$ prompts, each of which requires two inferences, the total time complexity of the entire stage is $O(2 \times n \times 1) = O(2n) = O(n)$. In the second stage of SBR model training, since the second stage only needs to parse the stored inference embedding into tensors, its added time complexity is $O(n)$. 
The time complexity of this phase is $O(SBR + n)$, and $O(SBR)$ usually lies between $[n,n^{2}]$, so the time complexity of the LLM4SBR mainly depends on the original SBR model. 
Therefore, the total time complexity of the LLMSBR framework is $O(LLM4SBR) = O(SBR + n) \approx O(SBR)$, which depends on the time complexity of SBR and the size of the dataset.

In summary, the added space and time costs are small compared to the huge performance gains of the LLM4SBR framework.

\begin{figure*}
    \centering
    \includegraphics[width=1.0\textwidth]{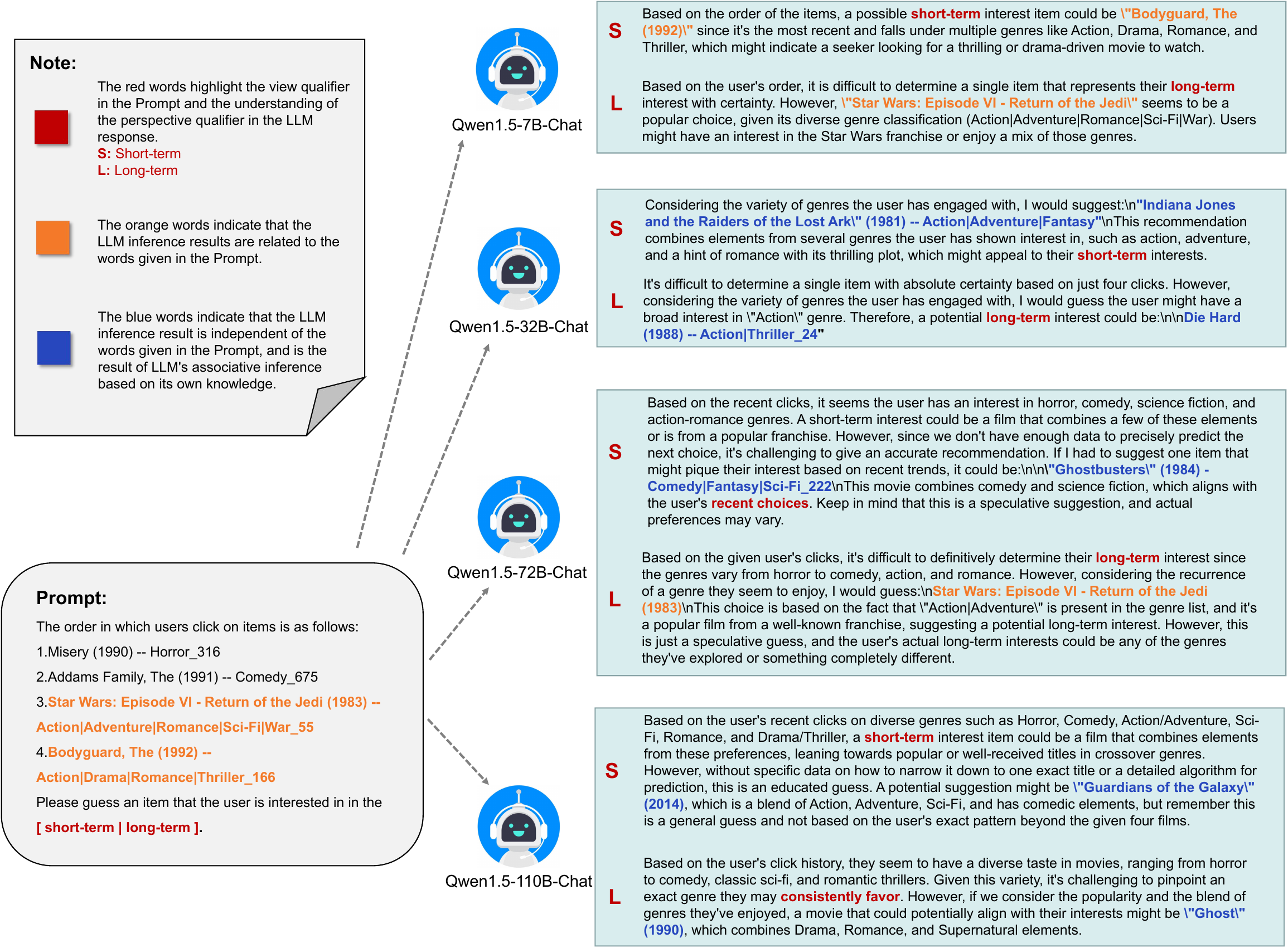}
    \caption{The inference results of LLM with different parameter quantities (7B, 32B, 72B, 110B) for short-term and long-term views.} 
    \label{fig:6}
\end{figure*}
\subsection{Case Study}
In this section, we analyze a specific case and examine the level of understanding and inference results of prompts by LLMs with different parameter quantities. 

As shown in Figure \ref{fig:6}, we generated prompts from both short-term and long-term views for the text of the same session sequence and used LLMs ( 7B, 32B, 72B, and 110B ) for inference. The results are shown on the right of the figure, where \textbf{"S"} represents the inference result of short-term interest and \textbf{"L"} represents the inference result of long-term interest. We did not remove the text explanations generated by LLMs here. We summarize the observed phenomena and analysis results as follows:
\begin{itemize}
    \item \textbf{All LLMs can respond to different view prompts, even those with the smallest parameter size.} In Figure \ref{fig:6}, we highlight in red the view-limiting qualifier in the prompt and the fields that respond to the view in the reply. We can see that all four LLMs in the figure understand the view-limiting qualifier in the prompt and make corresponding inferences in the response. In the inference results of the same LLM, the inference results of long-term interests and short-term interests are not the same. These phenomena indirectly prove that the inferencing ability of current general-purpose LLMs has been greatly improved and that they can infer reasonably within the original knowledge even without the fine-tuning of domain-specific knowledge.
    \item \textbf{The number of parameters in an LLM affects the inference ability and scope of a generalized LLM.} In the figure, we use orange to indicate that the inference item exists in the user sequence, and blue to indicate that the inference item does not exist in the user sequence. We can observe that LLMs of 7B tend to infer item representatives of users' long-term/short-term interests from the current sequence item set. In contrast, larger LLMs may generate inference results beyond the scope due to having more built-in knowledge. However, the LLM4SBR framework we proposed has already considered this situation and designed an intent localization module that can effectively constrain this LLM hallucination phenomenon. Furthermore, as shown in Table \ref{table 2}, even the 7B model can still bring significant performance improvements to the SBR model. The LLM4SBR framework allows free substitution of LLMs so that the appropriate LLM can be selected for inference depending on the actual requirements or computational resources, which provides a very high degree of flexibility and compatibility.
\end{itemize}

\section{Conclusions and Future work}
\label{sec::conclusion}

In this paper, we propose a scalable two-stage LLM enhancement framework (LLM4SBR) tailored for SBR. This approach is more efficient in utilizing information compared to encoding text data into embeddings for training, and it allows us to place LLM and SBR in separate stages, greatly reducing training costs. Specifically, in the semantic inference phase, we utilize LLM as the inference engine, designing prompt-guided inference processes from different views and leveraging an intent localization module to alleviate LLM hallucinations and enhance semantic. In the representation enhancement stage, we perform fine-grained alignment and uniformity of text embeddings and session embeddings from different views. This effectively facilitates the fusion of representations from different modalities, thereby enhancing the final session representation. Extensive experiments have demonstrated the effectiveness of the LLM4SBR framework, which significantly enhances most SBR models while also improving model interpretability and enhancing the diversity of candidate selection.

For future work, we will continue exploring whether adding additional LLM inference views can yield greater benefits, as well as assessing the effectiveness of utilizing LLM Agent for logical inference. In addition, we also want to explore the application of other downstream tasks combined with LLM. Finally, we hope for this work to open up new avenues in SBR research, accelerating deeper exploration into the integration of LLM with RS.

\begin{acks}
This work is supported by the National Key Research and Development Program of China under 2022YFB3104702.
This work is also supported by the National Natural Science Foundation of China under grants 62272262, 72442026, 72342032, 72074036, and 62072060. 
\end{acks}

\bibliographystyle{ACM-Reference-Format}
\bibliography{mybib}

\end{document}